\documentclass[%
reprint,
superscriptaddress,
amsmath,amssymb,
longbibliography,
aps,
nofootinbib,
prb
]{revtex4-2}
\usepackage[version=4]{mhchem}
\usepackage{graphicx}
\usepackage{amsmath,amssymb,amsthm,bm,hyperref,dcolumn}
\usepackage{comment}
\usepackage{xcolor}
\usepackage{pdfpages}
\usepackage{tikz}
\usetikzlibrary{arrows.meta}
\newcommand{\nmax}{n_\text{max}}
\newcommand{\lmax}{l_\text{max}}
\newcommand{\ylm}{Y^m_\ell}
\newcommand{\br}{\mathbf{r}}
\newcommand{\tbr}{\tilde{\mathbf{r}}}
\newcommand{\bx}{\mathbf{x}}
\newcommand{\Rhat}{\hat{R}}
\newcommand{\dref}{d_\mathrm{ref}}
\newcommand{\mcal}[1]{\ensuremath{\mathcal{#1}}}
\newcommand{\mbf}[1]{\ensuremath{\mathbf{#1}}}
\NewDocumentCommand{\glm}{}{\lambda\mu}

\newcommand{\bfeat}{\boldsymbol{\xi}}
\newcommand{\tbfeat}{\tilde{\boldsymbol{\xi}}}
\newcommand{\jac}{\boldsymbol{J}}

\NewDocumentCommand{\rep}{s d<| d|>}{%
\IfBooleanTF{#1}{
   \IfValueTF{#2}{
       \IfValueTF{#3}{\braket{#2}{#3}}{\bra{#2}}
       }{
       \IfValueTF{#3}{\ket{#3}}{}
       }
   }{
   \IfValueTF{#2}{
       \IfValueTF{#3}{\braket*{#2}{#3}}{\bra*{#2}}
       }{
       \IfValueTF{#3}{\ket*{#3}}{}
       }
   }
}

\NewDocumentCommand{\rbra}{sm}{\IfBooleanTF{#1}{\rep*<#2|}{\rep<#2|}}
\NewDocumentCommand{\rket}{sm}{\IfBooleanTF{#1}{\rep*|#2>}{\rep|#2>}}
\NewDocumentCommand{\rbraket}{smom}{
    \IfBooleanTF{#1}{
        \IfNoValueTF{#3}{\rep*<#2||#4>}{\rep*<#2|#3\rep*|#4>}
    }{
        \IfNoValueTF{#3}{\rep<#2||#4>}{\rep<#2|#3\rep|#4>}
    }
}

\NewDocumentCommand{\lm}{e_}{
\IfValueTF{#1}{l_{#1}m_{#1}}{lm}
}
\NewDocumentCommand{\nlm}{e_}{
\IfValueTF{#1}{n_{#1}\lm_{#1}}{n\lm}
}

\usepackage{xr}
\usepackage{xr-hyper}
\usepackage{hyperref}
\makeatletter
\AtBeginDocument{\let\LS@rot\@undefined}
\makeatother
\usepackage{caption}
\usepackage{subcaption}
\newsavebox{\fwmapbox}
\newsavebox{\invmapbox}
\AtBeginDocument{%
  \sbox{\fwmapbox}{%
    \tikz[baseline=-0.4ex]
      \draw[densely dotted, line width=0.55pt,
            -{Straight Barb[angle=60:2.4pt 3.2pt]}]
        (0,0) -- (1.5em,0);%
  }%
  \sbox{\invmapbox}{%
    \tikz[baseline=-0.4ex]
      \draw[line width=0.7pt, -{Stealth[length=2.2mm, width=1.7mm]}]
        (0,0) -- (1.5em,0);%
  }%
}
\DeclareRobustCommand{\fwmap}{\usebox{\fwmapbox}}
\DeclareRobustCommand{\invmap}{\usebox{\invmapbox}} 

\begin{document}
\title{Reconstructing local environments from concise atomistic representations} 

\author{Jigyasa Nigam}
\email{jnigam@mit.edu}
\affiliation{Research Laboratory of Electronics,
Massachusetts Institute of Technology, Cambridge, MA 02139, United States}

\author{Tuong Phung}
\affiliation{Department of Electrical Engineering and Computer Science,
Massachusetts Institute of Technology, Cambridge, MA 02139, United States}

\author{Ameya Daigavane}
\affiliation{Department of Electrical Engineering and Computer Science,
Massachusetts Institute of Technology, Cambridge, MA 02139, United States}

\author{Aria Mansouri Tehrani}
\affiliation{Research Laboratory of Electronics,
Massachusetts Institute of Technology, Cambridge, MA 02139, United States}

\author{Tess Smidt}
\email{tsmidt@mit.edu}
\affiliation{Research Laboratory of Electronics,
Massachusetts Institute of Technology, Cambridge, MA 02139, United States}
\affiliation{Department of Electrical Engineering and Computer Science,
Massachusetts Institute of Technology, Cambridge, MA 02139, United States}

\begin{abstract}
\noindent
Symmetry-based representations of local atomic structure, such as the power spectrum or bispectrum, are routinely used to characterize the structural diversity of datasets and as input features for atomistic machine learning.
Although these descriptors systematically incorporate increasingly complex geometric correlations, it remains unclear if a given feature can be mapped back to a discrete point cloud, whether such a reconstruction is unique, and how changes in the descriptor are reflected in the underlying atomic geometry. The choice and discretization of the radial and angular bases, as well as the high dimensionality of the resulting feature vectors -- which may contain hundreds or thousands of components -- make this interpretation even more challenging. 
\noindent
In this work, we investigate the inverse problem of recovering atomic structures from local \emph{invariant} descriptors. We show that accurate reconstructions can be obtained from remarkably compact descriptors of different correlation orders, each comprising only a few tens of features. Even representations that are formally incomplete or locally ill-conditioned can be inverted to accurate geometric reconstructions of atomic environments across molecular and material datasets. Our reconstruction framework provides a general algorithmic means of identifying approximate degeneracies of invariant descriptors and recovering distinct atomic environments that cannot be distinguished by a given representation. Finally, by reconstructing atomic configurations from descriptors, we examine how perturbations in invariant descriptors of different correlation orders translate into structural distortions.
\end{abstract}

\maketitle

\section{Introduction}
Machine learning (ML) methods have become a standard component of the atomistic simulation toolbox and are routinely used as surrogates for expensive quantum mechanical calculations or to extend simulations to much larger system sizes and time scales~\cite{unke2021machine, westermayr2021perspective, xia2025evolution}. 
Central to these approaches is the representation of atomic structures through descriptors (synonymously representations or features), i.e., mappings from atomic configurations to (typically high-dimensional) symmetry-adapted features that retain the structural information relevant for predicting physical properties~\cite{musi+21cr, allen2026optimal}. 
These descriptors can be constructed analytically or learned implicitly from data in end-to-end models, but in both cases, their effectiveness depends on how faithfully they encode the chemical diversity and local structural information relevant to describing physical observables. 
The design of descriptors, whether through principled mathematical formulations or through data-driven model architectures, has been guided by long-standing physical and chemical intuition about the electronic structure problem. 

The first among these is the notion that complex condensed matter and molecular systems can be understood through local decompositions of their structure. Locality, therefore, underlies a broad range of atomistic modeling approaches, including classical force fields, cluster expansions, and divide-and-conquer electronic structure methods~\cite{kohn96prl,prod-kohn05pnas,yang1991direct,yang1995density}.
A second guiding principle has been symmetry. Physical observables transform in well-defined ways under rigid translations, rotations of the structure, and permutations of identical atoms. For example, scalar quantities such as energies remain invariant under these operations. Atomistic representations and ML models, therefore, incorporate the symmetry of the target quantity as an inductive bias to improve data efficiency and generalization. This inductive bias has shaped the design of atomistic representations from early descriptions used in classical force fields and semi-empirical quantum chemistry methodologies (based on internal coordinates) ~\cite{jensen2017introduction} to modern architectures. 

Over the last two decades, substantial progress has been made in extracting richer atomic features directly from nuclear positions and chemical composition~\cite{musi+21cr}, including Behler–Parrinello neural networks~\cite{behl-parr07prl}, the smooth overlap of atomic positions (SOAP)~\cite{bart+13prb}, moment tensor potentials (MTP)~\cite{MTP} and permutation invariant polynomials (PIPs)~\cite{braa-bowm09irpc,xie-bowm10jctc}. These developments have been consolidated within the atomic cluster expansion (ACE) framework~\cite{drau19prb} (which generalizes cluster expansions from lattice models~\cite{walle2008complete} to atomic point clouds) and the atom-centered density correlation (ACDC) framework~\cite{nigam2020recursive}. These frameworks provide a linearly complete basis to expand the local environment in terms of two-body (one-neighbor), three-body (two-neighbor), and higher-order correlations, and consequently provide a systematically improvable set of hierarchical representations that can describe both invariant and tensorial quantities.  
These local descriptors have enabled the prediction of a wide range of molecular quantities, from energies and forces (interatomic potentials) to electronic response properties such as dipoles and polarizabilities~\cite{behl-parr07prl,bart+10prl,drau19prb,deringer2021gaussian, bochkarev2024graph}, and electronic structure observables including densities of states~\cite{mahmoud2020learning,fung2022physically}, electron densities~\cite{grisafi2018transferable, lou2026long}, and effective single-particle Hamiltonians~\cite{hegde2017machine, zhang2022equivariant, niga+22jcp}. In addition to such supervised learning tasks, they have been widely used to identify and classify atomic-scale structures~\cite{lechner2008accurate, rosenbrock2017discovering, dietz2017machine, hofgard2026learning}, and serve as powerful tools for visualizing structural diversity within datasets, interpreting model latent spaces, and identifying correlations between structural features and predicted properties~\cite{de2016comparing, rosenbrock2017discovering, cersonsky2021improving, glielmo2021unsupervised, nicholas2021visualization, donkor2023machine}. In nearly all of these applications, descriptors are forward mappings from atomic structures to feature vectors.

The inverse problem of recovering atomic structures from descriptors remains comparatively much less understood even though it is closely related to inverse design, which is hailed as the holy grail of materials and chemical discovery~\cite{zeni2025mattergen, merchant2023scaling}. This problem underlies several fundamental challenges in atomistic modeling, such as assessing the robustness of descriptors (whether constructed analytically or learned from data) and understanding which structural degrees of freedom are preserved, compressed, or effectively lost when transformed to a given local representation.

The problem of recovering structures satisfying prescribed constraints is not entirely new and has motivated several methods in atomistic modeling, from random structure search~\cite{pickard2011ab} for crystal structure prediction to approaches that reconstruct configurations from structural descriptors. In alloy theory, for example, special quasi-random structures (SQS)~\cite{zunger1990special, lian2025highly} are generated by searching over lattice configurations that reproduce a target SQS correlation function as closely as possible. Another popular approach formulates this task as a stochastic optimization problem over lattice occupations and generates configurations that match a set of correlation functions~\cite{walle2013efficient}.
Within atomistic ML, several works have examined the invertibility and completeness of local representations. Refs.~\citenum{bart+13prb, pozdnyakov2020completeness, pozdnyakov2021local} show that many commonly used invariant descriptors are \emph{incomplete}, such that geometrically distinct atomic environments can map onto identical descriptors. This kind of degeneracy had also been identified well over a decade earlier in Ref.~\citenum{boutin2004reconstructing}, which proposed configurations unrelated by rigid transformations that share distances, areas, and higher-order simplex volumes (which would be analogous to 2,3,4 body invariants\footnote{However, Ref.~\citenum{boutin2004reconstructing} compared global structural invariants instead of atom-centered correlations, so the analogy to hierarchical atom-centered descriptors is only approximate.}). In fact, several carefully parameterized families of distinct atomic environments with identical three- and four-body order invariants have been identified~\cite{pozdnyakov2020completeness}.
It remains unclear whether additional degeneracies exist beyond these known classes, particularly at higher body orders where completeness is expected but has not been concretely established~\cite{domina2026llm}, so approaches to building invariants that are provably complete and equivariant are being investigated~\cite{dusson2022atomic, kurlin2022computable, nigam2024completeness, maennel2024complete, kurlin2026scd}. 
Despite the non-injectivity of these representations, several works have demonstrated that atomic structures can, in practice, be recovered from local invariant descriptors such as atom-centered symmetry functions, power spectrum, and bispectrum~\cite{uhri21prb, fung2022atomic, cobelli2022local}. More recent generative approaches have used local many-body descriptors directly within structure generation pipelines~\cite{elijosius2025simgen}. Several works have also examined the use of descriptor-space search and optimization to reduce dataset complexity by identifying a set of representative atomistic configurations that reproduce target descriptor statistics corresponding to a larger ensemble~\cite{goff2025generalized}, or by using gradients of descriptor-based diversity metrics to generate and collate a compact but diverse training set~\cite{karabin2020entropy}.

In this work, we consider the challenge of recovering atomic environments from \emph{concise} invariant descriptors. We examine whether a unique atomic configuration (modulo symmetry operations) can be recovered from featurizations that are substantially more truncated than those typically used in atomistic ML, and investigate how the reconstruction accuracy depends on the basis resolution and correlation order. We note that when geometrically distinct environments cannot be distinguished by the representation, i.e., the descriptor is \emph{incomplete}, the reconstruction algorithm, starting from different initial random seeds, recovers a distribution of symmetry-inequivalent configurations that reproduce the descriptor equally well.
We further explore how interpolating between invariant descriptors translates to structural transformations, and conversely, how structural variation is reflected as changes in descriptor space. In doing so, we recover new pairs of geometrically distinct configurations that share nearly identical three- and four-neighbor correlation descriptors. 
We focus exclusively on mapping descriptors to local atomic structures, isolating it from the broader inverse design problem of identifying structures that exhibit prescribed target properties (which would involve an additional step of mapping from desired properties to descriptors). 

\begin{figure*}
    \centering
    \includegraphics[width=0.7\linewidth]{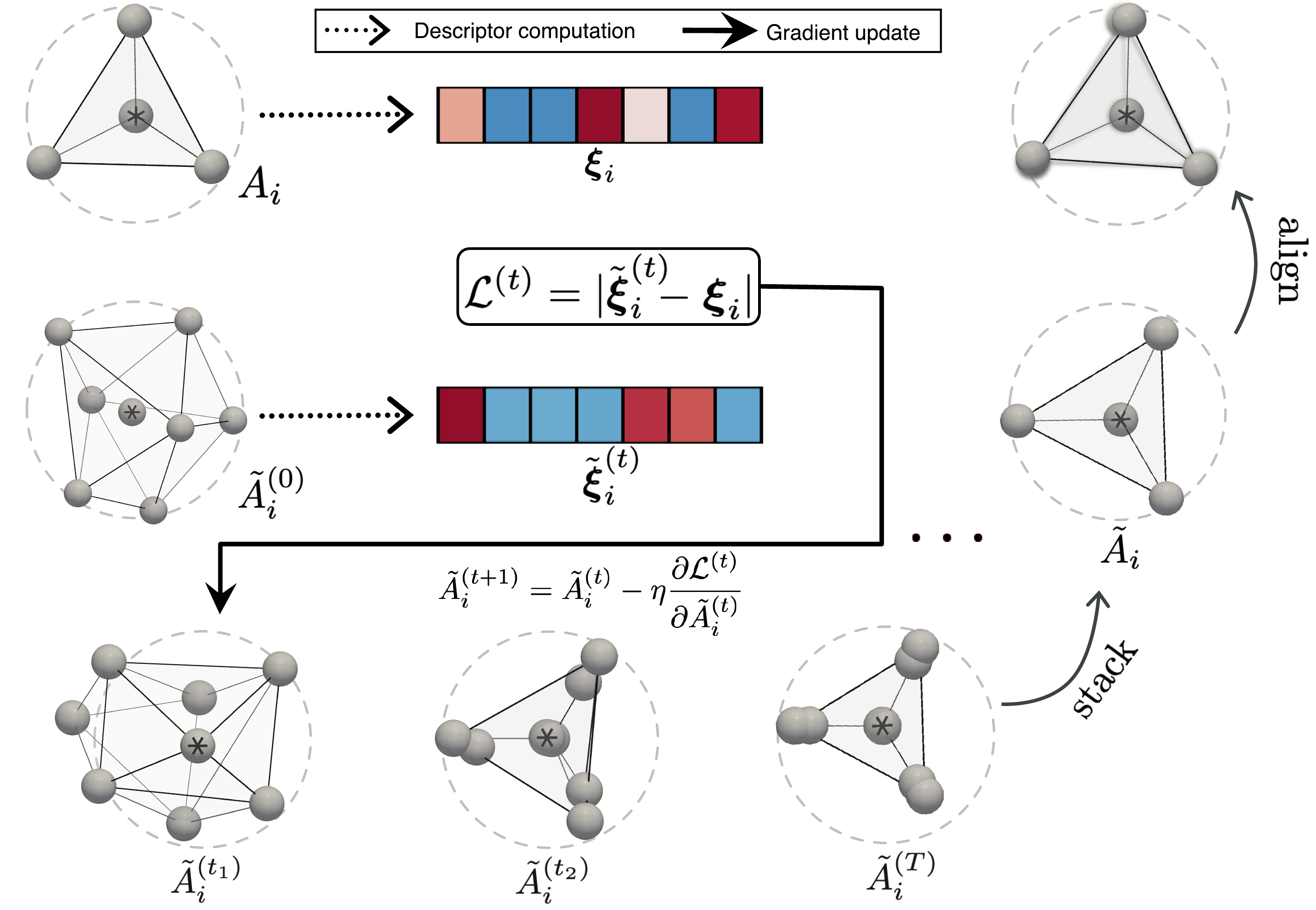}
    \caption{Mapping a reference local environment $A_i$, centered on atom $i$ (indicated with an asterisk) with $N_i$ neighboring atoms to an invariant descriptor $\bfeat_i$ (forward mapping in dotted lines \fwmap).
    To recover a structure that matches the descriptor, we begin with a randomly initialized trial environment  $\tilde{A}^{(0)}_i$ containing $N_i'$ atoms. For each optimization time step $t$, invariant descriptors are computed via the same forward mapping and compared to the target descriptor through a loss function $\mcal{L}^{(t)}$. The trial atomic coordinates are iteratively updated via gradient descent (inverse mapping in solid lines \invmap) to minimize the descriptor mismatch until a configuration with the desired number of atoms and a consistent descriptor is obtained. The configuration so recovered after $T$ optimization steps, $\tilde{A}^{(T)}_i$ (shown on the bottom right), can be aligned to the reference environment if desired.
    }
    \label{fig:schematic}
\end{figure*}

\section{Methodology}
To better understand the challenges of the inverse problem, we first review the forward map from an atomic configuration, $A$, specified by the set of nuclear coordinates $\br$ and atomic number $Z$ of each atom $i$, to atom-centered descriptors $\bfeat_i$.

\subsection{Mapping atomic configurations to invariant descriptors}
For each atom $i$ in $A$, we define a local spherical environment $A_i$ as the set of all neighboring atoms $j$ whose relative distance from the central $i$ satisfies $r_{ji} = \|\br_j-\br_i\| \le r_\text{cut}$, where $r_\text{cut}\in\mathbb{R}^+$ is the cutoff radius. Equivalently and more generally, the environment can be defined through a cutoff function $f_\text{cut}(r)$ which may be non-spherical. The construction above is recovered by choosing $f_\text{cut}(r)$ to be a spherical step function where $f_\text{cut}(r)=1$ for $r\le r_\text{cut}$ and $f_\text{cut}(r)=0$ otherwise. In practice, many atomistic ML models typically replace this step cutoff with a smooth function that decays continuously to zero at $r_\text{cut}$. 
We use the shorthand $A_i \equiv \{ \br_{ji}\}_{j=1}^{N_i}$ to denote the local environment of $i$, which is specified by the set of all interatomic vectors relative to $i$, $\br_{ji} = \br_j - \br_i$, and $N_i$ is the number of atoms within the cutoff. The local environment is represented by an atom-centered density,
\begin{equation}
\label{eq:atom-density}
\rho_i(\bx) = \sum_{j=1}^{N_i} \delta( \bx - \br_{ji})  \, f_\text{cut}(r_{ji}),
\end{equation}
where $\delta$ denotes the Dirac delta distribution. Following ACE~\cite{drau19prb}, we use the Dirac delta function $\delta$ to denote atomic positions, although other localized functions such as Gaussians are also widely used in other implementations~\cite{bart+13prb, will+19jcp, caro2019optimizing, witt2023acepotentials}.

This real-space description is then projected onto a finite basis of radial functions $R_n(r)$ (indexed by $n$) and angular functions, usually chosen to be (real) spherical harmonics (indexed by total $l$ and azimuthal $m$ indices). The radial and total angular indices are truncated at $\nmax$ and $\lmax$, respectively, while for each total angular index $l$, $m$ lies in the range $-l \leq m \leq l$. 
The resulting \emph{density coefficients} are given by
\begin{equation}
\label{eq:density-coeff-general}
\begin{aligned}
    c_{i}^{nlm} &= \int d\bx \, R_n(x) \ylm(\hat{\bx}) \rho_i (\bx) \\ & = \sum_{j \in A_i} R_n(r_{ji}) \, \ylm(\hat{\br}_{ji}) \, f_\text{cut}(r_{ji}) \\
    & = \sum_{j \in A_i} c_{ij}^{nlm}, 
\end{aligned}
\end{equation}
where we substituted Eq.~\eqref{eq:atom-density} and used the Dirac delta to limit the sum to atoms $j \in A_i$ in the local environment.

The density coefficients can be extended to incorporate the dependence on the neighbor chemical species through an additional `alchemical' basis $Z$, by restricting the sum over neighbors corresponding to the species $Z$ as,
\begin{equation}
\label{eq:density-coeff}
c_{i}^{Znlm} = \sum_{\substack{j \in A_i \\ Z_j =Z}} c_{ij}^{nlm}.
\end{equation}
In modern architectures, such as graph neural networks, this alchemical basis is often replaced by learnable atomic embeddings that encode the chemical nature of each atom. 
Since the focus of this work is to probe the geometric relationship between atomic environments and invariant descriptors, we omit the alchemical basis in the following.

As each density coefficient $c_{i}^{nlm}$ is obtained by summing contributions of the relative orientation of individual neighbors, it corresponds to a one-neighbor (two-body, including the central atom) correlation and does not describe the interaction between neighbors.  
To incorporate this information, the density coefficients can be combined to form higher-order features. For instance, a $\nu$-neighbor (or $\nu+1$ body) correlation centered on $i$ obtained through a tensor product of $\nu$ copies of the density coefficients,
\begin{equation}
\label{eq:nu-order-unsymm}
  \rho_i^{\otimes \nu} = \prod_{\alpha=1}^\nu c_{i}^{n_\alpha l_\alpha m_\alpha},
\end{equation}
when explicitly expanded, decomposes into a sum of $\nu$-tuples of neighbors $(j_1, \ldots, j_\nu)$
\begin{equation}
    \prod_{\alpha=1}^\nu c_{i}^{n_\alpha l_\alpha m_\alpha}
    = \prod_{\alpha=1}^\nu \left( \sum_{j_\alpha} c_{ij_\alpha}^{n_\alpha l_\alpha m_\alpha} \right)
    = \sum_{j_1, \dots, j_\nu} \prod_{\alpha=1}^\nu c_{ij_\alpha}^{n_\alpha l_\alpha m_\alpha}.
\end{equation}
To obtain symmetry-adapted descriptors, the tensor product is followed by symmetrization over the O(3) group, 
\begin{equation} 
\label{eq:nu-order-symm}
\bfeat_{i; \mbf{n}\mbf{l}}^{\otimes \nu; \glm ;\sigma } = \sum_{\mbf{m}} \rho_i^{\otimes \nu} \mbf{C}^{\glm}_{\mbf{l}\mbf{m}},
\end{equation}
where $\mbf{n} \equiv (n_1, \ldots, n_\nu)$ denotes the tuple of radial channels, $\mbf{l} \equiv (l_1, \ldots, l_\nu)$ the tuple of total angular momenta for each copy of the density, and $\mbf{m} \equiv (m_1, \ldots, m_\nu)$ is the tuple of azimuthal angular momenta. $\mbf{C}^{\glm}_{\mbf{l}\mbf{m}}$ is the generalized Clebsch-Gordan (CG) coefficient that combines $\nu$ angular momenta $ \mbf{l}\mbf{m}\equiv (\lm_{1}, \ldots, \lm_{\nu}$) to total angular momentum $\glm$.
The additional superscript $\sigma$ denotes the parity under inversion, and is computed as $\sigma = (-1)^{l_1 + l_2 .. + l_\nu + \lambda}$. It takes the value $-1$ or 1 if the resulting symmetrized feature behaves as a pseudo or polar tensor under inversion, respectively. For ease of notation, we will denote descriptors invariant under (im)proper rotations corresponding to $\lambda = \mu=0, \sigma = 1$, using the shorthand $\bfeat_i^{\otimes \nu, 1}$ or more compactly as $\bfeat_i^{\otimes \nu}$ when it is clear from context. Similarly, pseudoscalars are denoted as  $\bfeat_i^{\otimes \nu, -1}$. It is also standard to refer to descriptors corresponding to $\nu=2,3,4$ as the power spectrum, bispectrum, and trispectrum, respectively. This procedure of mapping local environments to descriptors is shown as dotted lines in Fig.~\ref{fig:schematic}.

In Eq.~\eqref{eq:nu-order-symm}, $\mbf{n}\mbf{l}$ enumerate the distinct radial and angular channels contributing to the given $\bfeat_i^{\otimes \nu}$ and determine the feature dimensionality for a fixed basis truncation $\nmax$ and $\lmax$. 
The number of features grows rapidly with correlation order and radial resolution, scaling exponentially as $\nmax^\nu$. The dependence on $\lmax$ is less straightforward, as allowed combinations of $\mathbf{l}$ are restricted by CG coefficients. The resulting number of features and the computational scaling vary with how this selection is implemented algorithmically~\cite{xie2025price, dusson2022atomic}. For instance, not all $\mathbf{l}$ tuples need to be treated independently, since permutations of a given tuple correspond to different coupling paths within the same tensor product space and are related. N$-$body iterative contraction of equivariants (NICE) ~\cite{nigam2020recursive} reduces this redundancy by retaining unique $\mbf{l}$ tuples obtained through a canonical ordering of angular indices $l_1 \leq l_2 \leq \cdots \leq l_\nu$. Reduced tensor products (RTP) in e3nn~\cite{geiger2022e3nn} implement a complementary solution by constraining the symmetrization from $\rho^{\otimes \nu}$ to $\bfeat$ to be simultaneously symmetric under SO(3) and permutations of the angular indices $(l_\alpha m_\alpha)$. Gaunt tensor products~\cite{luo2024enabling}, on the other hand, bypass explicit CG symmetrization by evaluating angular couplings through integrals of spherical harmonics on the $S^2$ grid (surface of a three-dimensional sphere). Further details on how these formulations are related are provided in Section~\ref{sisec:comparing-descriptor-implementation} of the Supplementary Information (SI).
Although these approaches differ in how the features are computed, the recovered invariants are equivalent up to scaling. The only exception is that pseudoscalars ($\lambda=0, \sigma=-1$) cannot be recovered with Gaunt tensor products as they are restricted to describing polar spherical harmonic signals. 
For the remainder of this work, we present results using the NICE implementation (identical results are obtained with RTP), and note that the proposed inversion applies equally to all the formulations described above.

To mitigate the exponential scaling with radial bases, we also consider a simplified case of Eq.~\eqref{eq:density-coeff-general} with a single radial basis function $R_n(r) = r$. In this limit, the density coefficients take the form, 
\begin{equation}
\label{eq:density-coeff-single}
c_{i}^{lm} = \int d\bx \, \ylm(\hat{\bx}) \rho_i (\bx) \, = \sum_{j \in A_i} r_{ji} \, \ylm(\hat{\br}_{ji}) \, f_\text{cut}(r_{ji}) .
\end{equation}

As our goal is not predictive accuracy but the inversion process itself, this greatly reduces the size and computational cost of the feature vectors and disentangles the inversion dynamics from a specific form or parametrization of the radial basis. We note that compactness need not necessarily come at the expense of accuracy, as property-aware feature compression has recently been shown to improve both the computational efficiency and predictive accuracy~\cite{banjafar2026property} of descriptors. 

The representation resulting from our property-agnostic basis truncation in Eq.~\eqref{eq:density-coeff-single}, however, effectively resolves structure only through a single contribution along each angular direction. For a fixed direction $\hat{\mathbf{u}}$,
\begin{equation}
\label{eq:simplified-density-coeff}
    c_{i}^{lm} =  \sum_{j_{\hat{\mbf{u}}}} r_{ji}\, f_{\text{cut}}(r_{ji})\, Y_l^m(\hat{\mathbf{u}})   = Y_l^m(\hat{\mathbf{u}}) \, F_{i}(\hat{\mbf{u}}),
\end{equation}
where $ j_{\hat{\mbf{u}}}$ indicates the collinear neighboring atoms lying along $\hat{\mbf{u}}$ and $F_{i}(\hat{\mbf{u}}) = \sum_{j_{\hat{\mbf{u}}}} r_{ji}\, f_{\text{cut}}(r_{ji})$ is the aggregated radial factor. 
This can introduce additional degeneracies during inversion (as collinear atoms collapse to a single coefficient), as discussed in SI Section~\ref{sisec:radial}. These degeneracies can be systematically removed by increasing the radial resolution or imposing additional constraints on the environment. One such constraint is the total number of particles in a given local environment, which is recovered either as the $\nu=0$ (one-body) invariant~\cite{drau19prb}, or equivalently, from the invariant $l=0$ term in Eq.~\eqref{eq:density-coeff-general} when using a constant radial basis $R_n(r) = 1$,
\begin{equation}
   \bfeat_i^{\otimes 0} \equiv  \sum_{j \in A_i} Y_{0}^{0}(\hat{\br}_{ji}) \propto N_i
\end{equation}
which simply counts the number of atoms in the environment.
Alternatively, the cutoff function $f_\text{cut}(r)$ may be chosen such that only neighboring atoms within the first coordination shell contribute to Eq.~\eqref{eq:simplified-density-coeff} to avoid projecting multiple collinear atoms onto the same density coefficient.

\subsection{Inverting descriptors to atomic environments}
We now formulate the inverse problem (shown as solid lines in Fig.~\ref{fig:schematic}). Given a target atom-centered descriptor $\bfeat_i$ corresponding to an environment $A_i$, we seek a local environment $\tilde{A}_i$, specified by the set of relative positions $\{\tilde{\br}_{ji}\}$ of neighboring atoms, whose descriptor $\tilde{\bfeat}_i$ reproduces the reference as closely as possible.

In practice, recovering atomic environments from \emph{invariant} atom-centered descriptors is challenging for several reasons. First, since configurations related by rigid rotations, inversions, and permutations of identical atoms are intentionally mapped to the same descriptor, any representative within this structurally equivalent set constitutes a valid reconstruction. Second, as noted above, a combination of truncated angular or radial basis resolution and finite correlation order can accidentally map distinct environments to identical descriptors~\cite{pozdnyakov2020completeness,nigam2024completeness}. 
As a result, inverting the descriptor becomes equivalent to recovering a distribution of configurations. Configurations related through encoded rotation or permutation symmetries (structural degeneracies) can be identified by aligning reconstructed environments to a fixed reference, as described in SI Section~\ref{sisec:alignment}. Accidental degeneracies, on the other hand, are substantially more difficult to characterize, and identifying them generally requires extensive sampling of configuration space.

This global non-bijectivity is further complicated by the local ill-conditioning of the inverse map. Following Ref.~\citenum{pozdnyakov2021local}, this can be understood from the Jacobian of the forward map,
\begin{equation}
\label{eq:jac}
    \jac = \frac{\partial \bfeat_i}{\partial {\br_{ji}}},
\end{equation}
which quantifies how perturbations of atomic coordinates affect the descriptor. An infinitesimal perturbation of atomic coordinates $\delta \br$ results in the corresponding first-order change in the descriptor of
\begin{equation}
\label{eq:jacobian-lin}
    \delta \bfeat_i \approx  \jac \delta \br.
\end{equation}

Small singular values of $J$ correspond to directions in configuration space for which comparatively large coordinate displacements result in weak changes in the descriptor. Equivalently, due to this instability of the inverse mapping, two feature vectors with a small difference can correspond to substantially different atomic geometries.
Combined with the nonlinear (polynomial) dependence of atom-centered descriptors on interatomic displacement vectors, these effects produce a highly non-convex reconstruction landscape with multiple local minima.  
Nevertheless, the differentiability of the forward map from atomic configurations to descriptors allows gradients with respect to atomic coordinates to be computed efficiently, making it possible to optimize the local environment. We formulate inversion as a minimization of the reconstruction loss $\mcal{L}(\tbfeat_i(\{\tbr_{ji}\}), \bfeat_i(A_i))$ (e.g.\ mean squared error or mean absolute error), where the trial descriptor $\tilde{\bfeat}_i$ is computed from a trial atomic environment $\{\tbr_{ji}\}$. The loss is optimized iteratively over the atomic positions in the neighborhood through gradient descent,
\begin{equation}
\label{eq:optimization}
    \tbr_{ji}^{(t+1)} = \tbr_{ji}^{(t)} - \eta
\frac{\partial \mcal{L}^{(t)}}{\partial \tbr_{ji}^{(t)}},
\end{equation}
where $\eta$ denotes the learning rate, $\mcal{L}^{(t)}$  the optimization loss and $\tbr_{ji}^{(t)}$ the position of atom $j$ relative to the central atom $i$ at optimization step $t$.

\begin{figure*}
    \centering
    \includegraphics[width=1.0\linewidth]{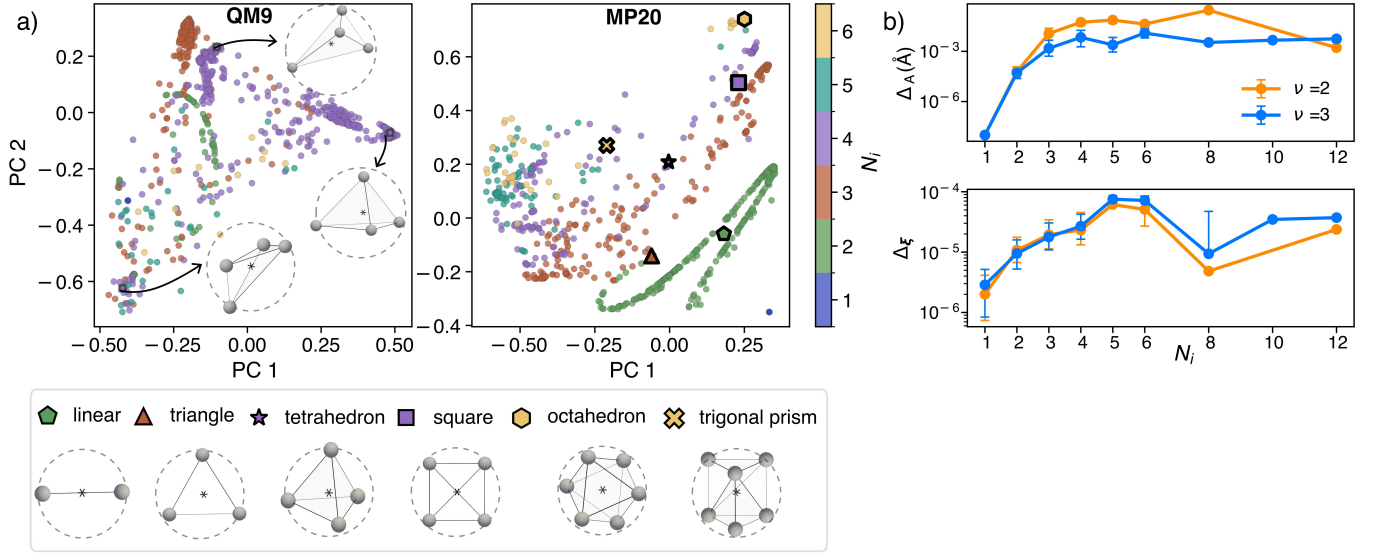}
    \caption{a) PCA projection of the bispectrum ($\nu = 3$) descriptors computed for local environments extracted from QM9 and MP20 datasets, colored by the number of neighbors, $N_i$, in the reference environments. To facilitate interpretation, we also mark a few highly symmetric environments (linear, triangle, tetrahedron, square, octahedron, and trigonal prism) on the second panel and show the environments as insets in the legend.
    b) Descriptor reconstruction error ($\Delta_{\bfeat}$) and structural reconstruction error (RMSD, $\Delta_{A}$) as a function of coordination number for the power spectrum ($\nu=2$) in orange and bispectrum ($\nu=3$) in blue for successful reconstructions. Error bars denote the standard deviation across environments with the same $N_i$ across three independent reconstruction attempts.}
    \label{fig:qm9-mp}
\end{figure*}

The initial configuration for optimization, $\tilde{A}^{(0)} = \{\tbr_{ji}^{(0)}\}$, need not be initialized with the same number of atoms as the reference environment. Instead, we begin with a larger number of initial points, treating it as a hyperparameter, and allow the configuration to relax as the loss is optimized. The effective number of neighboring atoms in the environment is refined by collapsing points whose angular separation relative to the central atom falls below a threshold $\theta_{\text{thresh}}$, as shown in Fig.~\ref{fig:schematic}. Additional points may also be introduced if merging reduces the number of points below the desired number of neighbors. In the following, we will denote by $\tilde{A}_i$ the environment resulting from $T$ optimization steps followed by collapsing the neighbors and, if desired, alignment with a reference environment. 

In some of the experiments reported below, we use multiple random initializations to explore the non-convex landscape and increase the likelihood of recovering a point cloud with the desired descriptor, which is also physically plausible (for example, maintains a desired minimum interatomic separation). 

\section{Results}
In the following discussion, we use the mean absolute error between the predicted and reference descriptors as the optimization loss $\mcal{L}$. To quantify descriptor reconstruction accuracy, we report the root mean square error (RMSE) of the reconstructed descriptors as $\Delta_{\bfeat}$,
\begin{equation}
    \label{eq:rmse}
    \Delta_{\bfeat} = \sqrt{\frac{1}{N_\text{test}} \sum_{i=1}^{N_\text{test}} \lvert \tilde{\bfeat}_i - \bfeat_i \rvert^2},
\end{equation}
where $\bfeat_i $ is the target descriptor and $\tbfeat_i$ is the descriptor corresponding to the optimized environment. We additionally compute the root mean square deviation (RMSD), $\Delta_A$, between the reconstructed environment $\tilde{A}_i$ and the reference environment $A_i$,
\begin{equation}
    \label{eq:rmsd}
    \Delta_A = \sqrt{\frac{1}{N_\text{test}} \sum_{i=1}^{N_\text{test}} \min_{\Rhat}\, \lvert \Rhat\tilde{A}_i - A_i \rvert^2},
\end{equation}
as a measure of the structure reconstruction accuracy. Since the descriptors are invariant under rotations, RMSD is computed after aligning $\tilde{A}_i$ to the reference using $\Rhat$ as described in SI Section~\ref{sisec:alignment}.

As a complementary metric we also compute $\dref$ introduced in Ref.~\citenum{bart+13prb},
\begin{equation}
    \dref = \min_\mbf{P} \, \lVert \mbf{D} - \mbf{P} \tilde{\mbf{D}} \mbf{P}^T \rVert
\end{equation}
where $\mbf{P}$ is the permutation matrix, $\mbf{D}$ and $\tilde{\mbf{D}}$ are the distance matrices (defined in SI Section~\ref{sisec:alignment}) corresponding to the original environment $A_i$ and recovered environments $\tilde{A}_i$ respectively. As elements of the distance matrix depend only on pairwise distances, $\dref$ is invariant under global rotations and avoids the need for an explicit rotation alignment, making this metric particularly useful when the optimal alignment becomes difficult to determine due to reconstruction errors or noise. However, one must optimize the permutation of atom labels,  $\mbf{P}$, to find the best match of atoms in the original and recovered structures.
For all of the following experiments, the differentiable forward mapping from environments to descriptors and these metrics were implemented in PyTorch~\cite{paszke2019pytorch}.

\subsection{Recovering environments in molecular and material datasets}
\label{sec:results-qm9mp20}
We begin by evaluating the extent to which local atomic environments can be reconstructed from invariant descriptors across 1000 atomic environments from the Materials Project Dataset~\cite{jain2013materials} and the QM9 dataset~\cite{ramakrishnan2014quantum} each. We supplement these datasets with 730 environments from transition metal complexes~\cite{balcells2020tmqm} (taken from Ref.~\citenum{atomic-datasets}) and high symmetry configurations (triangle, hexagonal bipyramid, hexagonal prism, cube, octahedron, pentagonal prism, and an icosahedron) containing up to 12 atoms.
For each structure in these datasets, we extract a representative atomic environment using a CrystalNN cutoff function implemented in PyMatgen~\cite{ong2013pymatgen}. This cutoff retains the nearest neighbors through Voronoi decomposition of the local environment, and reduces the descriptor degeneracies arising from collinear neighboring atoms as described above.

For $\nu=2$, we compute invariants with $\lmax = 8$, resulting in a compact feature vector of size 9. To control the growth of feature size with increasing correlation order, we use $\lmax = 6$ for $\nu=3$. Note that invariant features computed at $\nu=2$ are necessarily even under inversion and therefore do not contain pseudoscalars irrespective of $\lmax$. For $\nu=3$, the first pseudoscalar contribution arises from coupling density coefficients corresponding to the angular tuple $\mbf{l} = (2,3,4)$, requiring a minimum $\lmax$ of 4 to allow reflection-sensitive information to be included in reconstruction (see Fig.~\ref{sifig:tetrahedra} for a comparison of reconstructions with and without pseudoscalars). 
The resulting $\nu=3$ descriptors combining 30 scalars and 5 pseudoscalars are vectors of size 35.

Fig.~\ref{fig:qm9-mp}a shows a two-dimensional projection of the principal component analysis (PCA) of $\nu=3$ descriptors for the QM9 and MP20 datasets. As the descriptors do not include alchemical information, the clustering reflects purely geometric variations of the local environments. For instance, the most prominent trend is a clear stratification of descriptors according to the number of neighbors (degree of ``coordination" of the central atom), reflecting the diversity of local environments present in the datasets, from singly-coordinated atoms to environments containing more than ten neighbors. Representative symmetric environments (triangular, square, tetrahedral, \ldots) are marked on the right panel to ease interpretation. Within each cluster corresponding to a particular number of neighbors, the descriptor further resolves additional structural differences. For example, environments with four neighbors in the QM9 dataset form several distinct clusters, corresponding to different geometric arrangements, including nearly regular tetrahedral, distorted tetrahedral, and bent nearly planar environments (shown as insets in the left panel of Fig.~\ref{fig:qm9-mp}a).   
Thus, even very compact invariant descriptors retain sufficient information to distinguish subtle structural differences between local atomic environments. We assess this more rigorously by quantifying the accuracy with which local environments can be reconstructed from their featurizations.

We characterize reconstructions with a mean absolute error loss $\mcal{L}< 10^{-4}$ to be successful, resulting in 67\% successful reconstructions from $\nu=2$ and 90\% successful reconstructions from $\nu=3$. Fig.~\ref{fig:qm9-mp}b) shows the RMSDs and the mean absolute error of the optimized descriptors of the recovered environments as a function of the number of neighbors in the reference local environment.
In principle, reconstructing atomic environments with $N_i$ neighbors up to global rotations and translations requires fixing $3 N_i -3$ degrees of freedom (DOF) (translation invariance accounted for by fixing the central atom at the origin). A necessary condition for $\bfeat$ to be injective is then that its dimensionality $d$ satisfy $d \geq 3 N_i -3$. However, our results show that accurate reconstructions can be obtained even when $d < 3 N_i -3$, as is the case for $\nu=2$ (where $d=9$) for most environments, including $N_i$ up to  12, possibly due to the symmetries/structural constraints that further reduce the degrees of freedom in the environment. 
In addition to dimensionality (or correlation order), the angular resolution ($\lmax$) seems to be critical for accurate reconstructions. For example, a higher correlation order \(\nu=4\) descriptor with 56 components but limited angular resolution $\lmax=3$ has a much lower success rate of reconstruction. (See SI Section~\ref{sisec:molecular-data} for a comparison with reconstructions from 56-dimensional $\nu=4$ features.) 
\begin{figure}
    \centering
    \includegraphics[width=1.\linewidth]{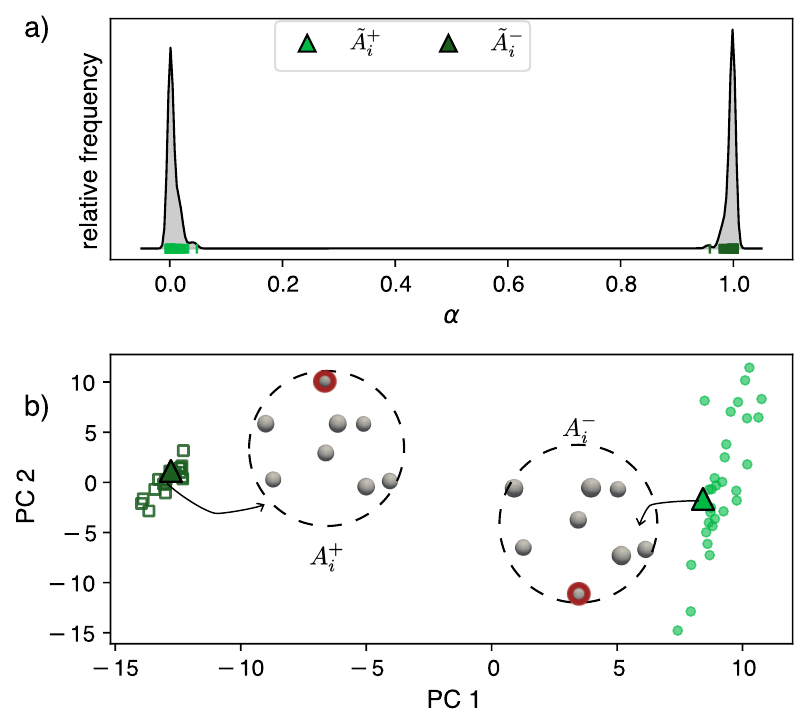}
    \caption{a) Distribution of successful reconstructions across five bispectrum-degenerate \ce{B8} pairs and 100 trials for each pair. Recovered environments with a root mean squared error greater than $1 \times 10^{-5}$ were discarded, resulting in a total of 169 successful inversions. Each resulting environment is binned according to the coordinate $\alpha = d_+ / (d_+ + d_-) $, where $d_+$ and $d_-$ denote its trispectrum distance to the corresponding reference $A_i^+$ and $A_i^-$. The resulting distribution is sharply bimodal, and reconstructions cluster near $\alpha \approx 0$ or $\alpha \approx 1$, showing that the reconstruction recovers both reference structures.
    b) PCA projection of $\nu=4$ features of the recovered environments for one of the five pairs of degenerate environments. The reconstructed configurations separate cleanly into two clusters around the two distinct \ce{B8} clusters (corresponding local environments shown in the inset). }
    \label{fig:pca-b8}
\end{figure}

\begin{figure}
    \centering
    \includegraphics[width=1.0\linewidth]{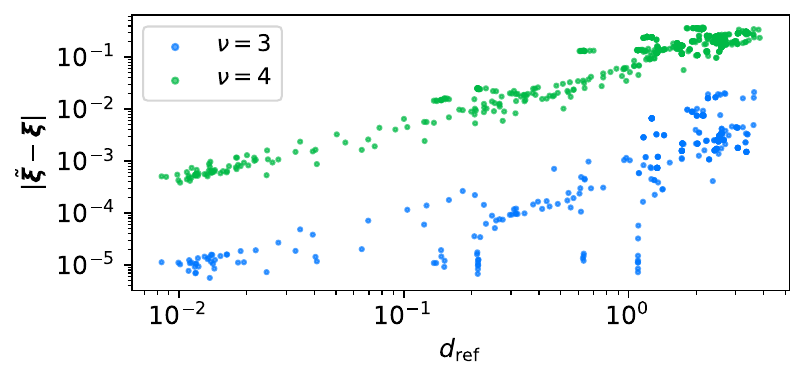}
    \caption{For each recovered geometry, we plot its geometric deviation $d_\mathrm{ref}$ against the corresponding descriptor mismatch $|\tilde{\boldsymbol{\xi}}^{\otimes \nu} - \boldsymbol{\xi}^{\otimes \nu}(A_i^\pm)|$ (the difference between the descriptor of the recovered geometry and that of the corresponding reference environment $A_i^+$ and $A_i^-$). Bispectrum ($\nu=3$) is shown in blue and trispectrum ($\nu=4$) in green. A small descriptor mismatch means the recovered geometry is nearly indistinguishable from the reference in feature space, whereas a large $\dref$ means it is nonetheless geometrically different from that reference. The vertical bands in the bispectrum indicate that geometries spanning a wide, non-negligible range of $\dref$ all map to essentially the same descriptor. By contrast, the trispectrum mismatch grows smoothly with the geometric separation between the recovered and reference structures, so a small trispectrum mismatch implies the recovered structure is geometrically close to the reference as well.
   }
    \label{fig:sensitivity-b8}
\end{figure}

\subsection{Inverting degenerate descriptors}
We turn to examples of inverting accidentally degenerate descriptors, which are known to be shared by pairs of geometrically distinct atomic environments. These experiments serve two purposes: first, to verify if we recover both the reference environments associated with the descriptor, and second, to explore whether additional environments beyond the known pair can reproduce the same bispectrum.
We consider 5 random pairs of \ce{B8} clusters reported in Ref.~\cite{zenodo-three-center} constructed following the procedure detailed in Refs.~\cite{pozdnyakov2020completeness, nigam2024completeness}. In these constructions, the central atom is fixed at the origin, and the resulting environments, $A_i^+$ and $A_i^-$, differing only by the position of a single neighbor atom (as highlighted in the insets of Fig~\ref{fig:pca-b8}b), share identical $\nu=3$ correlations.

Starting from 100 random initializations of the atomic coordinates, we perform independent reconstructions by minimizing the loss on descriptors, $\mcal{L}$, for each target descriptor. A reconstruction is considered successful if the final descriptor optimization loss (RMSE) satisfies $\mathcal{L} < 10^{-5}$, ensuring that the recovered structure reproduces the target descriptor to numerical precision. Across all successful reconstruction attempts, the average descriptor error $\Delta_{\bfeat}$ for the bispectrum was $4.7 \times 10^{-6}$.

To assess whether the inversion procedure recovers distinct geometric configurations, we compute a higher-order $\nu=4$ descriptor (trispectrum) for the reconstructed environments, which resolves the degeneracy between $A_i^+$ and $A_i^-$. 
For each recovered structure $\tilde{A}_i$, we compute the difference of its trispectrum and those of the two reference degenerate structures for the corresponding trial
\[
d_{\pm}(\tilde{A}_i) = \lVert\boldsymbol{\xi}^{\otimes 4}(\tilde{A}_i) -
\boldsymbol{\xi}^{\otimes 4}(A_i^\pm)\rVert,
\]
and define a normalized coordinate
$\alpha = \frac{d_+}{d_+ + d_-} $ restricted between 0 and 1. Values of $\alpha$ closer to zero indicate that the recovered environment is closer to $A_i^{+}$, while $\alpha$ closer to one means the recovered environment is closer to $A_i^{-}$. Fig.~\ref{fig:pca-b8}a) shows the distribution of $\alpha$ across all successful reconstructions, which separates into two peaks around $\alpha=0$ and $\alpha =1$, demonstrating that the inversion of the degenerate bispectrum recovers one of the two parameterized symmetry-inequivalent geometries. Fig.~\ref{fig:pca-b8}b) shows a PCA projection of the trispectrum of the resulting successfully reconstructed structures for one degenerate pair. The recovered structures cluster around two distinct basins corresponding to the two symmetry-inequivalent environments $A_i^+$ and $A_i^-$, consistent with accidental degeneracy in the descriptor. The RMSD $ \Delta_A$ between the successfully recovered structures and the corresponding reference structure is $3.2 \times 10^{-4}$~\AA{}, confirming an almost exact reconstruction.

We have focused so far exclusively on exact reconstructions, but in practice, the reconstruction algorithm often converges to errors that are small yet above the success threshold. This leads us to ask whether descriptors close to each other map onto geometrically similar structures, analogous to the question examined in Ref.~\citenum{widdowson2026pointwise}, albeit for a different family of invariants. If structures associated with nearly identical feature values are also geometrically similar, the representation is stable and sensitive to local structural distortions. Conversely, if small descriptor errors correspond to large structural deviations, then some geometric degrees of freedom are only weakly constrained by the descriptor (as will be discussed in the next section). To quantify this behavior, we compute for each ``failed" reconstructed structure (i.e., those where the RMSE between the target and recovered descriptors $>10^{-5}$)  the geometric distance $\dref$ to both its reference configurations $A_i^+$ and $A_i^-$. Fig.~\ref{fig:sensitivity-b8} compares this geometric deviation with the corresponding difference in both $\nu=3$ and $\nu=4$ descriptors.
Despite large geometric deviations from the reference structure, many reconstructions remain extremely close with respect to the bispectrum, which is an artifact of the degeneracy in the descriptor being optimized (and for this reason, a low descriptor reconstruction error need not necessarily imply the recovered geometry is close to either reference). In contrast, the trispectrum distance increases smoothly with geometric separation from the reference environments. The higher-order descriptor, therefore, resolves geometric variations that remain essentially indistinguishable by the bispectrum. 
\begin{figure}
    \centering
    \includegraphics[width=1.\linewidth]{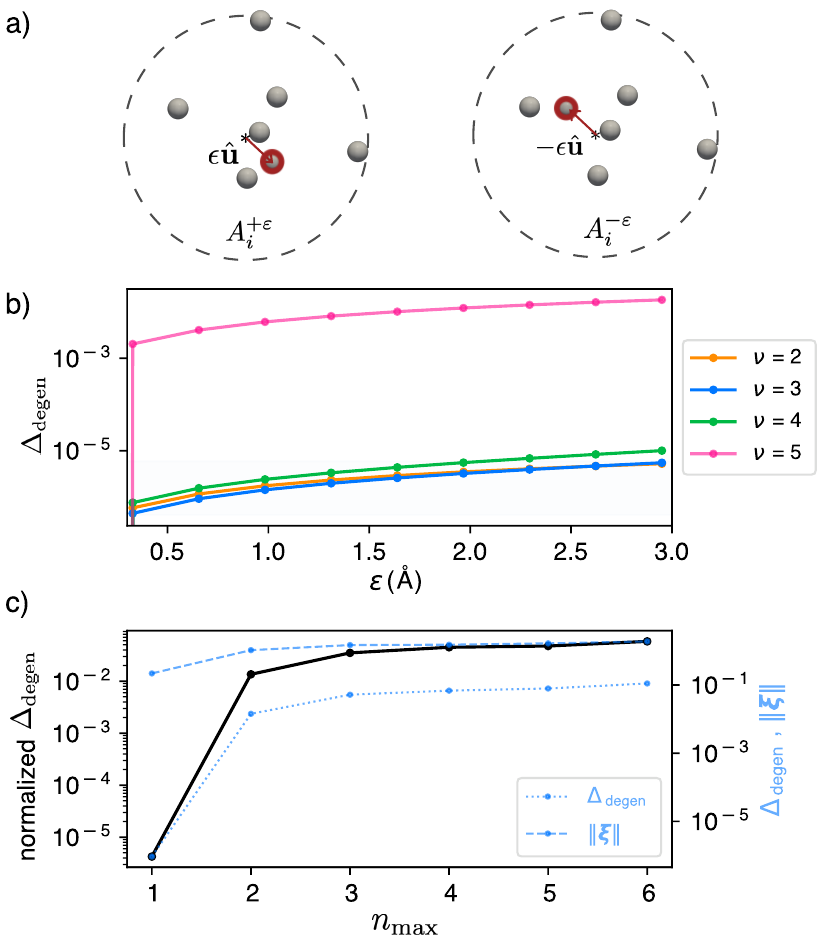}
    \caption{a) A pair of distinct environments containing 7 neighboring atoms with almost degenerate $\nu=4$ descriptors. Starting from a reference environment $A_i^{(0)}$, an additional atom (highlighted in red) is displaced by $\pm \varepsilon \hat{\mbf{u}}$ along an optimized direction $\hat{\mbf{u}}$ to generate the two configurations $A_i^{+\varepsilon}$ and  $A_i^{-\varepsilon}$. Despite the descriptors being nearly identical ($\Delta_\text{degen} \approx 10^{-6}$), the geometric deviation $\dref$ between the structures is 2.1~\AA{}.  The atomic coordinates for these environments are reported in the SI. 
    b) The near-degeneracy between $A_i^{+\varepsilon}$ and  $A_i^{-\varepsilon}$, quantified by $\Delta_\text{degen}$, persists as the additional atom is displaced from $\varepsilon= 0$~\AA{} to 3 ~\AA{} along $\hat{\mbf{u}}$ in both positive and negative directions. The resulting structures remain nearly indistinguishable up to $\nu=4$, but are distinguished by $\nu=5$ descriptors.
    c) Since the optimization of atomic coordinates is performed with a single radial basis $(\nmax = 1)$, increasing the basis resolution $\nmax$ partially resolves the near degeneracy. $\Delta_\text{degen}$ (dotted blue, right axis) grows along with the descriptor norm $\|\bfeat\|$ (dashed blue, right axis). Their ratio, normalized $\Delta_\text{degen}$ (black, left axis), also increases with $\nmax$. }
    \label{fig:new-degen}
\end{figure}

\subsection{Identifying nearly accidentally-degenerate structures}
\label{sec:accidental-degen}
The observation that certain geometric distortions can produce negligible changes in the descriptor motivates a constructive strategy for identifying distinct atomic environments that are nearly degenerate for a given descriptor. Mathematically, such directions can be characterized through local conditioning of the descriptor with respect to atomic displacements and expressed through the Jacobian $\jac$. As discussed in Ref.~\cite{pozdnyakov2021local}, the singular value decomposition of the Jacobian defined in Eq.~\eqref{eq:jac},
\begin{equation}
\jac = \mbf{U} \mbf{\Sigma} \mbf{V}^\top,
\end{equation}
provides a linearized description of feature sensitivity. The right singular vectors (columns of $\mbf{V}$) associated with small singular values define displacement modes in Eq.~\eqref{eq:jacobian-lin} along which the descriptor changes weakly, i.e., $\delta \boldsymbol{\xi}_i \approx 0$.

Rather than directly evaluate the spectrum of the Jacobian, we consider a finite-difference proxy of the same local sensitivity. Starting from an initial atomic environment $A_i^{(0)}$, comprising $N_i$ neighbor atoms around the origin, we introduce a fictitious atom placed at a displacement of magnitude $\varepsilon$ along a direction vector $\hat{\mathbf{u}}$ relative to the origin. We evaluate the sensitivity of the descriptor to displacements along this direction through the finite difference,
\begin{equation}
\Delta_{\mathrm{degen}} =
\left\|
\boldsymbol{\xi}(A_i^{+\varepsilon}) -
\boldsymbol{\xi}(A_i^{-\varepsilon})
\right\|.
\end{equation}
where $A_i^{(\pm \varepsilon)}$ denotes the configuration in which the additional atom is displaced along $\pm \varepsilon \, \hat{\mathbf{u}}$, respectively.

\begin{figure*}
    \centering
    \includegraphics[width=0.7\linewidth]{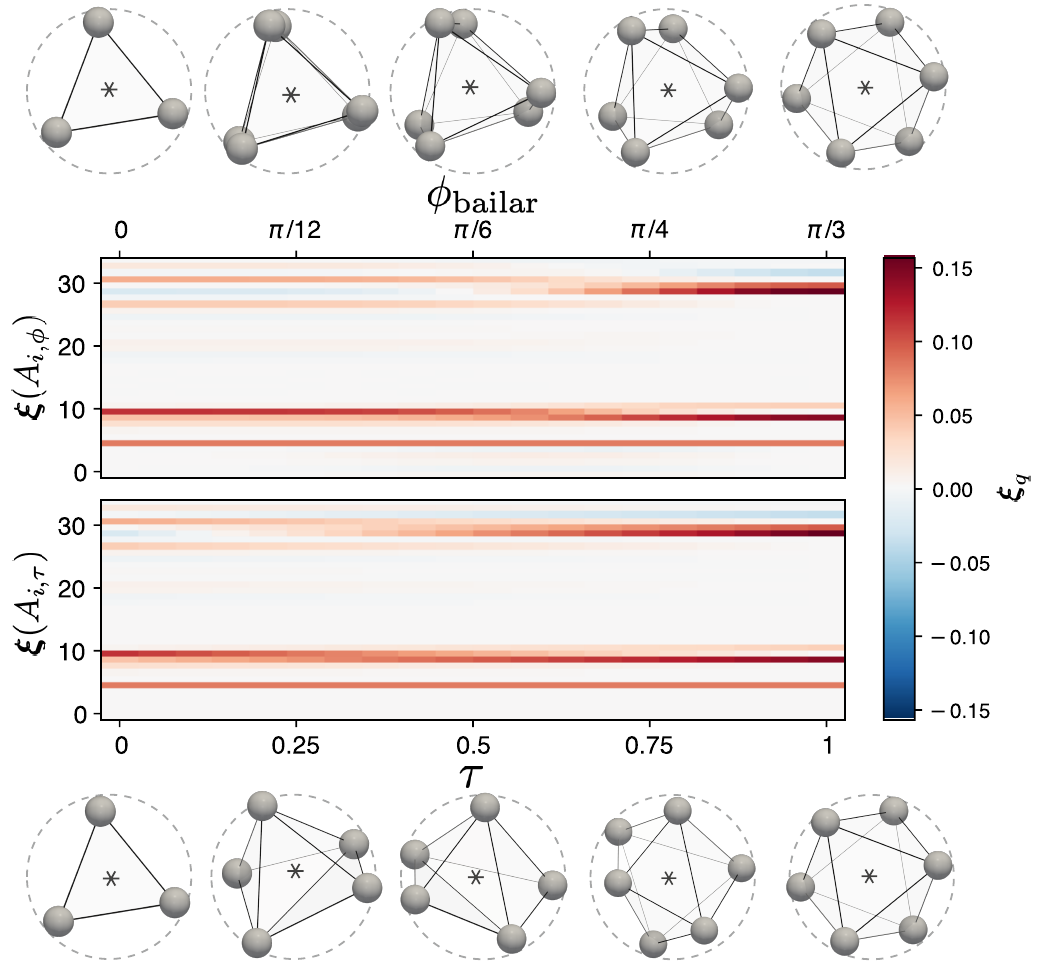}
    \caption{Comparison between the bispectrum ($\nu=3$) descriptors along the Bailar twist ($\bfeat(A_{i,\phi})$), parameterized by the twist angle $\phi_\text{bailar}$ and those obtained by linearly interpolating between descriptors corresponding to the endpoints, parameterized by $\tau \in[0,1]$ ($\bfeat_{i,\tau}$) over twenty steps each. The descriptors were computed with $\lmax = 6$ and include both $\sigma=1$ and $\sigma=-1$ components, resulting in a 35-dimensional vector, colored by the values of the descriptor component indexed by $q$. 
    Reference environments $A_{i, \phi}$ sampled along the twist from trigonal prism ($\phi_\text{bailar}=0$ and octahedron $\phi_\text{bailar}=\pi/3$) are shown at the top, and point clouds $\tilde{A}_{i, \tau}$ reconstructed by inverting the interpolated descriptor are shown at the bottom. 
    Although the two descriptor trajectories look qualitatively similar (middle panels), the reconstructed structures show that linear interpolation in descriptor space introduces geometric distortions that are absent from the true Bailar transformation. See SI Fig.~\ref{sifig:desc-component-bailar} for more details.}
    \label{fig:interpolate-bailar}
\end{figure*}

By minimizing $\Delta_{\mathrm{degen}}$ with respect to both the coordinates of the initial atomic configuration $A_i^{(0)}$, $\varepsilon$, and the direction $\hat{\mathbf{u}}$ (coordinates of the additional atom), we recover configurations comprising $N_i+1$ atoms with nearly identical values of the descriptor. To ensure physical plausibility of the resulting point cloud, we include an additional geometric regularization that penalizes solutions with overlapping atoms or very small interatomic distances. We include an additional penalty on $\dref$ during optimization to favor accidental degeneracies over trivial structural degeneracies (that is, to avoid recovering environments related by rotations or atom-relabeling, which would minimize the loss to zero by design). 

Using this constructive approach, we recover geometrically distinct environments that share nearly identical $\nu=4$ descriptors, reducing $\Delta_{\mathrm{degen}}$ to $1.2 \times 10^{-6}$ while maintaining $\dref$ of 2.1~\AA{}. A representative example containing seven neighbors is shown in Fig.~\ref{fig:new-degen}a). 
(We also report a pair of near-degenerate configurations at $\nu=3$ in the SI). Although these examples correspond to near rather than exact degeneracies, they show that substantial geometric differences can map onto a small $\Delta_{\mathrm{degen}}$, so predictive models for quantities that are sensitive to these changes cannot be represented as smooth functions of the descriptor. 
Since $\nu=4$ contains all lower body-order descriptors as its subset, these optimized environments are also degenerate at $\nu=2$ and $\nu=3$. Increasing the correlation order, however, resolves the degeneracy, increasing $\Delta_{\mathrm{degen}}$ by two to three orders of magnitude as shown in Fig.~\ref{fig:new-degen}b). Continuing to distort the structures along the optimized direction preserves the near-degeneracy, indicating that the degeneracy extends beyond this single optimized pair of structures, but forms an extended, weakly constrained direction in configuration space. 

Unlike known analytical constructions, which are restricted to a specific coordination number or parameterization of neighbors, our approach provides a general numerical procedure for identifying degenerate configurations at arbitrary body order. 
As the optimization for degenerate coordinates was performed using descriptors with a single radial basis, it is partially lifted as the radial resolution increases, as illustrated in Fig.~\ref{fig:new-degen}c). A similar optimization can be performed with richer radial bases. However, as the descriptor dimensionality grows exponentially with body order and radial resolution, searching for near-degenerate environments becomes computationally more demanding, and the optimization may depend on the choice of the radial basis. 

\subsection{Interpolation between diverse atomic environments}
\begin{figure}
    \centering
    \includegraphics[width=1.0\linewidth]{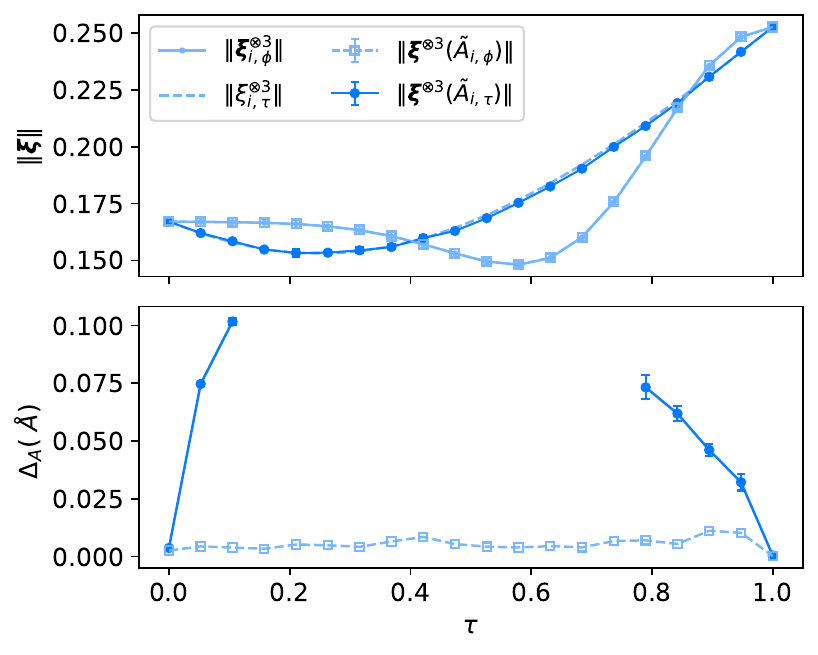}
    \caption{Descriptor norms (top) and the structural reconstruction error ($\Delta_A$) (bottom) along the Bailar twist path for $\nu=3$, plotted against the interpolation coordinate $\tau$ (when referencing the twist, we use $\tau$ to mean the normalized twist angle $\phi/ (\pi/3)$). Solid light blue lines show the norm of the reference descriptor trajectory $\bfeat(A_{i,\phi})$ (the geometric twist), dashed light blue lines show the norm of the linearly interpolated descriptors $\bfeat_{i,\tau}$. Filled circles (dark) and open squares show the norms of descriptors computed from the corresponding reconstructed environments $\tilde{A}_{i, \tau}$ (optimized to match $\bfeat_{i,\tau}$) and $\tilde{A}_{i, \phi}$ (optimized to match $\bfeat(A_{i,\phi})$), respectively. When the reconstruction optimization achieves zero loss, the filled circles coincide with the dashed line and the open squares coincide with the solid lines; deviations between markers and their corresponding lines therefore indicate imperfect reconstruction. 
    In the bottom panel, we show the $\Delta_A$ between the reference structures $A_{i, \phi}$ and the two reconstructed structure trajectories ($\tilde{A}_{i, \tau}$ and $\tilde{A}_{i, \phi}$). Since $\Delta_A$ is undefined when the reference and reconstructed structures have different numbers of atoms, we use missing points in the bottom panel to denote cases where the reconstructed environment converged to a different number of neighbors than the reference. 
    }
    \label{fig:rmsd-interpolate-bailar}
\end{figure}

Finally, extending the preceding local sensitivity analysis, we examine whether different paths through descriptor space connecting the same endpoints correspond to different structural transformations. Given a descriptor trajectory associated with a continuous geometric deformation, we construct its simplest alternative by linearly interpolating between the descriptors at the endpoints and compare the environments reconstructed along both paths. 

We first consider the transformation from a trigonal prism to an octahedral coordination environment, each comprising six neighbors around a central atom. This transition, analogous to the Bailar twist, is particularly interesting as it can be described by a single order parameter, namely the twist angle ($\phi_\text{bailar}$) between the two triangular faces of the polyhedron. For simplicity, in the following discussion we denote $\phi_\text{bailar}$ as $\phi$. We discretize the twist into twenty intermediate environments $\{A_{i,{\phi}}\}$, where $\phi\in[0,\pi/3]$ parameterizes the continuous geometric deformation between the trigonal prism ($A_{i,{0}}$) and octahedron ($A_{i,{\pi/3}}$), and compute the corresponding descriptor trajectory denoted by $\{\bfeat(A_{i, {\phi}})\}$. 

Since the mapping from atomic coordinates to descriptors is nonlinear, a simple structural deformation does not necessarily correspond to a simple trajectory in descriptor space, nor does a simple continuous path in descriptor space necessarily correspond to a simple continuous geometric deformation. To test this correspondence, we construct a reference path in descriptor space by linearly interpolating between descriptors corresponding to the endpoints $\bfeat(A_{i,0})$ and $\bfeat(A_{i,\pi/3})$ over $n_\text{interpolate}$ steps, 
\begin{equation}
\label{eq:linear-interpolate}
    \bfeat_{i, \tau} = (1-\tau) \,\bfeat(A_{i,0}) + \tau \,\bfeat(A_{i, \pi/3}),
\end{equation}
where $\tau \in [0,1]$ is the discrete interpolation parameter, with $\tau=0$ corresponding to $\phi=0$ ( trigonal prism) and $\tau=1$ corresponding to $\phi = \pi/3$ (octahedron). Note that we denote the interpolated descriptor as $\bfeat_{i, \tau}$ (rather than $\bfeat(A_{i, \tau}$) since it is not guaranteed that a corresponding point cloud $A_{i, \tau}$ exists.
The deviation between the descriptors generated by the physical deformation, $\{\bfeat(A_{i, {\phi}})\}$, and linear interpolation, $\{ \bfeat_{i, \tau}\}$, quantifies the nonlinearity of the descriptor mapping along the Bailar twist.

To translate these deviations back into structural distortions, we invert each interpolated descriptor $ \bfeat_{i, \tau}$ to a local environment $\tilde{A}_{i,\tau}$. We then compare the resulting structural trajectory $\{ \tilde{A}_{i,\tau}\}$ with the Bailar twist $\{A_{i, \phi}\}$, to test whether it can also be described by a simple order parameter or if it contains distortions beyond the relative twist. 

To facilitate comparison, we set $n_\text{interpolate}=20$ (matching the number of intermediate structures used to discretize the geometric twist) throughout this analysis. 
Fig.~\ref{fig:interpolate-bailar} shows a comparison of the 35 components of the $\nu=3, \lmax=6$ descriptors corresponding to the Bailar twist, ${\bfeat(A_{i,\phi})}$, against those obtained from the linear descriptor interpolation ${\bfeat_{i,\tau}}$. The descriptor norms are compared in the top panel of Fig.~\ref{fig:rmsd-interpolate-bailar}. As expected, the two descriptor trajectories coincide at the endpoints $\tau=0$ and $\tau=1$ but diverge substantially (with a difference $\|\bfeat(A_i\phi) - \bfeat_{i, \tau}\|$ of $\approx 10^{-1}$)at the intermediate points. The difference of the two descriptor trajectories, together with the descriptor component exhibiting the largest deviation, is shown in SI Fig.~\ref{sifig:desc-component-bailar}. 

Inverting the interpolated descriptors $\bfeat_{i,\tau}$ results in environments $\tilde{A}_{i,\tau}$ that are significantly different from $A_{i,\phi}$, except at the endpoints. See the insets in Fig.~\ref{fig:interpolate-bailar} for five representative environments sampled from $\{A_{i, \phi}\}$ (top) and  $\{\tilde{A}_{i, \tau}\}$ (bottom). For intermediate values of $\tau$ (where the feature reconstruction accuracy is $\Delta_{\bfeat} \sim 3 \times 10^{-3}$, compared to $\Delta_{\bfeat} \sim 1 \times 10^{-5}$ at $\tau=0$ and $\tau=1$) the reconstructed environments $\tilde{A}_{i,\tau}$ do not necessarily preserve six-fold coordination, and the optimization often converges to structures with only four or five neighbors. Even when the recovered $\tilde{A}_{i,\tau}$ have the expected number of neighbors (closer to $\tau=0$ and $\tau=1$), the resulting environments have a high RMSD (up to 0.1 ~\AA{}) (Fig.~\ref{fig:rmsd-interpolate-bailar} bottom) over five independent reconstruction trials, as the reconstructed environments exhibit significant in-plane stretching and angular distortions. Thus, even though the linear interpolation in $\bfeat^{\otimes \nu}$ is continuous, the resulting structural evolution is neither smooth nor confined to the original geometric transition. In SI Fig.~\ref{sifig:rmsd-n6-bailar} we repeat this analysis while constraining the reconstruction along the interpolation trajectory to environments with six atoms. Although this constrained reconstruction achieves a comparable optimization loss, the reconstructed environments remain similarly distorted relative to $A_{i, \phi}$.

\begin{figure*}[]
    \centering
    \includegraphics[width=0.7\linewidth]{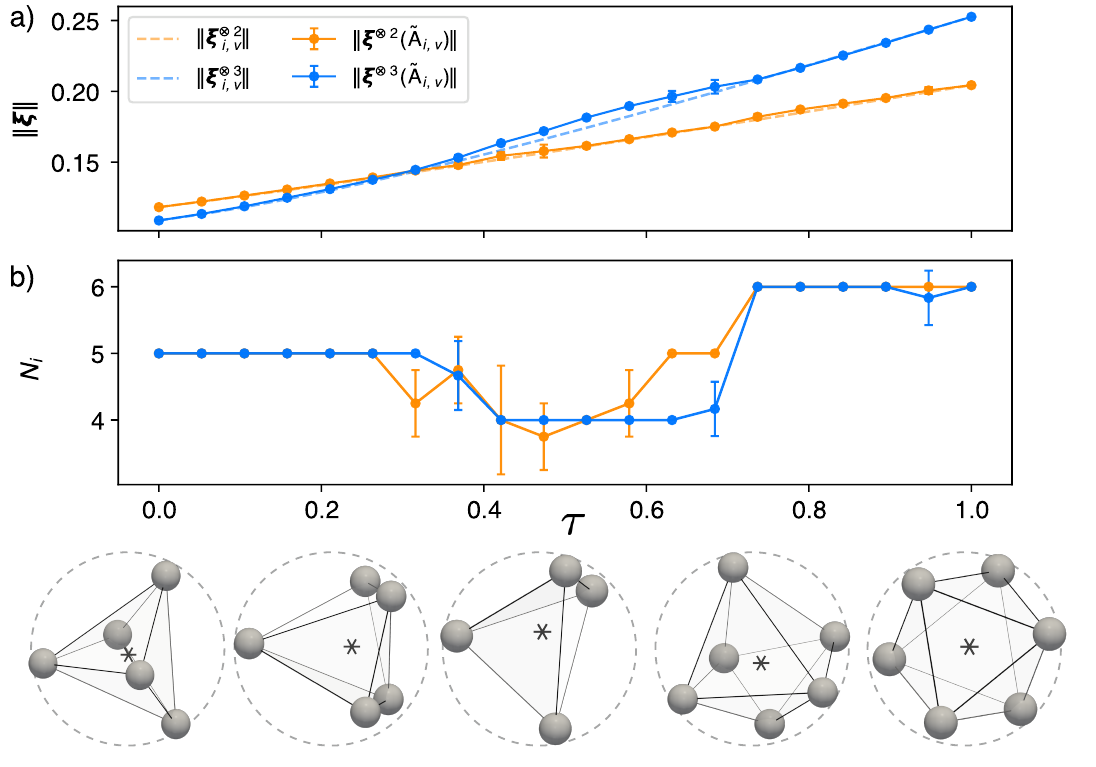}
    \caption{Linear interpolation of descriptors ($\nu=2$ in orange, $\nu=3$ in blue) corresponding to a trigonal pyramid environment (containing five neighbors) and an octahedral environment (containing six neighbors). a) Dashed lines show the norms of the linearly interpolated descriptors $\| \bfeat_{i, \tau}\|$. Filled circles (connected by solid lines) show the norm of the descriptor computed from the reconstructed environment $\| \bfeat(\tilde{A}_{i, \tau})\|$. The close agreement at both $\nu=2$ and $\nu=3$ confirms that reconstruction optimization recovers structures whose descriptors match the interpolated target, despite the change in coordination number between endpoints. 
    b) Number of neighbors in the recovered environment $\tilde{A}_{i, \tau}$ at each interpolation step $\tau$ for $\nu=2$ (orange) and $\nu=3$ (blue). While the endpoints $\tau=0$ and $\tau=1$ reconstruct accurately to the expected high-symmetry environments with five or six neighbors, descriptors at intermediate interpolation points consistently recover environments with around four neighbors. Representative reconstructed geometries at five interpolation steps are shown at the bottom. }
    \label{fig:interpolate-natoms}
\end{figure*}

As a second example, we consider transformations between local environments with different numbers of neighbors. Although these environments differ in the number of neighbors in the environment, they are represented as descriptors of fixed dimensionality, making interpolation in descriptor space considerably simpler than in configuration space. We consider the transformation between trigonal bipyramidal and octahedral coordination environments, which are frequently encountered motifs in coordination chemistry. 
By linearly interpolating between the two endpoints following Eq.~\eqref{eq:linear-interpolate} over twenty intermediate steps, we obtain a sequence of target descriptors, from which we then recover point clouds. 

Since the descriptor does not explicitly encode the number of neighbors, the inverse reconstruction must infer the coordination from the geometric information encoded in $\bfeat$. For each interpolated descriptor $\bfeat_{i,\tau}$, we perform five independent reconstruction trials. Fig.~\ref{fig:interpolate-natoms}a) shows that the norms of the reconstructed descriptors (solid lines) for both $\nu=2$ and $\nu=3$ match the corresponding target descriptor (dashed lines) closely throughout the interpolation trajectory. At the two endpoints $\tau=0$ (trigonal bipyramidal) and $\tau=1$ (octahedral), the lowest reconstruction error is achieved by environments containing five and six neighbors, as expected. However, as seen in Fig.~\ref{fig:interpolate-natoms}b), the coordination number does not vary smoothly along the interpolation. The continuous path in descriptor space maps onto discontinuous changes in the number of neighbors, and intermediate descriptors frequently are best inverted to environments that need not have the same number of neighbors as either endpoint. 

In both our examples, inverting descriptor trajectories to atomic structures exposes the structural consequences of seemingly simple operations in descriptor space. Our observations indicate that different continuous paths in descriptor space can map to qualitatively different structural transformations, including distinct distortions or discontinuous changes in the number of neighbors.

\section{Discussion and conclusion}
Local invariant descriptors have become standard geometric fingerprints of atomic structure. They are routinely used as \emph{succinct} representations of structural variations within materials and molecular datasets, and as inputs to data-driven models of atomistic properties. Although they encode geometric quantities (such as sets of interatomic distances, angles, dihedrals), it is not always clear whether a given descriptor uniquely determines an atomic environment, how sensitive the representations are to perturbations in the underlying environment, and how changes in descriptor space are reflected in geometric distortions.
Of these, the question of uniqueness and invertibility, in particular, has proven to be more subtle than it may appear. Ref.~\citenum{pozdnyakov2020completeness}, for example, identified distinct configurations that share $\nu=2$ and $\nu=3$, thus demonstrating that these descriptors are not globally invertible, i.e., a given set of invariants does not always correspond to a single atomic environment. Consequently, reconstructing atomic environments from these atom-centered descriptors is an intrinsically ill-posed problem. 

In this work, we examined descriptor invertibility and sensitivity in an even more challenging setting by adopting an extremely parsimonious discretization of the atomic density and using only a single radial basis function. Although this coarse discretization introduces additional degeneracies, it also compresses the descriptor to only a few tens of components, making it ideal to probe how much geometric information can still be recovered from invariant descriptors, and how this recovery depends on the choice of descriptor hyperparameters. 
Despite this aggressive truncation, we showed that local atomic environments can still be reconstructed with remarkably high accuracy across realistic molecular and materials datasets. Moreover, when accidental degeneracies are present, multiple independent inversion trials recover the distinct symmetry-inequivalent configurations compatible with the target descriptor and provide a numerical route to identifying incompleteness.
In addition to identifying global incompleteness, our reconstruction framework provides a constructive means of probing the local conditioning of invariant descriptors. 
We numerically identified geometric distortions along which the descriptor is insensitive by optimizing pairs of atomic environments that are nearly indistinguishable by $\nu=3$ and $\nu=4$ descriptors. Although these examples correspond to near rather than exact degeneracies, they provide promising candidates for future analytical investigation and can be systematically refined by performing similar optimization with a finer radial resolution.

The compactness of these descriptors also makes the relationship between configuration and descriptor spaces more interpretable. We examined how trajectories through descriptor space map onto trajectories in configuration space and found that some smooth changes in descriptors can lead to discontinuous changes in the recovered geometry (such as a change in coordination number of the central atom). 

Finally, throughout this work, we focused exclusively on reconstructing local atomic environments, deliberately separating it from the problem of recovering entire atomic structures and from the broader challenge of inverse design. This separation allowed us to identify the ingredients necessary for successful reconstruction and interpret changes in descriptors in terms of local structural distortions. Future work could extend this framework to reconstructing multiple overlapping environments or exploring descriptor space guided by its relationship to target properties before reconstructing candidate atomic environments. 

\begin{acknowledgments}
JN is grateful for funding from the MIT Postdoctoral Fellowship for Excellence in Engineering. JN and TS acknowledge funding from the Air Force Office of Scientific Research under Award No. FA9550-24-1-0067 and the National Science Foundation under Cooperative Agreement PHY-2019786 (The NSF AI Institute for Artificial Intelligence and Fundamental Interactions).
TP acknowledges funding from the Thomas and Stacey Siebel Foundation, Analog Devices, the MIT Undergraduate Research Opportunities Program, and the Air Force Office of Scientific Research for their support through Grant FOA-AFRL-AFOSR-2023-0011. AD was supported by the NSF Graduate Research Fellowship program under Grant No. DGE-1745302.

\noindent
\textbf{Conflict of interest statement}
The authors have no competing interests to declare.\\
\textbf{Data availability} All the code used for this study will be made publicly available upon publication.
\\
\textbf{Author contributions}
TS conceived the project and, with JN, designed the overall research methodology. JN and TP developed the algorithmic and code infrastructure, with contributions from AD. JN wrote the first draft of the manuscript. All authors discussed the results and reviewed the final version of the manuscript.

\end{acknowledgments}

\bibliography{refs}
\onecolumngrid\clearpage


\section*{Supplementary material}

# Reconstructing local environments from concise atomistic representations Supplementary Information

Jigyasa Nigam,[1,*] Tuong Phung,[2] Ameya Daigavane,[2] Aria Mansouri Tehrani,[1] and Tess Smidt[1,2,†]

[1]*Research Laboratory of Electronics, Massachusetts Institute of Technology, Cambridge, MA 02139, United States*
[2]*Department of Electrical Engineering and Computer Science, Massachusetts Institute of Technology, Cambridge, MA 02139, United States*

## I. OPTIMIZING POINT CLOUDS VS DENSITY EXPANSION COEFFICIENTS

The invariant descriptor $\boldsymbol{\xi}$ is computed as polynomials of the expansion coefficients of the atomic density. One could then formulate the reconstruction problem in two independent steps, by first recovering the density coefficients and subsequently inverting those coefficients to atomic configurations. The second step, however, is non-trivial as the recovered coefficients must correspond to a discrete point cloud rather than an arbitrary density. In general, a set of density expansion coefficients that generate a target descriptor need not always correspond to a discrete atomic configuration.

To see why this is the case, recall that spherical harmonics form a smooth, complete basis for functions on the sphere, meaning any square-integrable function $f(\theta,\phi)$ can be written as

$$f(\theta,\phi)=\sum_{l=0}^{\infty}\sum_{m=-l}^{l} c_{lm}Y_l^m(\theta,\phi). \tag{S1}$$

If two functions $f_1$, $f_2$ have identical coefficients $c_{lm}$ in Eq. (S1), for *all* $l$ and $m$'s, then they are identical. However, when the expansion is truncated at $l_{\max}$, distinct functions can share the same low-order expansion coefficients and appear to be identical. As a result, infinitely many distinct functions can share the same truncated expansion. As a simple example, consider a function specified by the sum of Dirac delta functions located at the vertices of a regular tetrahedron inscribed within a sphere of radius $R$,

$$\rho(\mathbf{x})=\sum_{i=1}^{4}\delta(\mathbf{x}-\mathbf{r}_i),$$

where $\mathbf{r}_i$ are the positions of the vertices. Due to the tetrahedral symmetry, only spherical harmonic coefficients corresponding to $l=0,3,4,6,\ldots$ (i.e., those consistent with tetrahedral symmetry) are non-zero. For a density expansion truncated at $l_{\max}=4$, it is possible to construct smooth functions that identically reproduce these coefficients (but differ in higher-angular coefficients). Consider the one-parameter family of smooth densities on the surface of the sphere,

$$\rho(\mathbf{x})=\delta(x-R)\left[\sum_{l\in\{0,3,4\}}\sum_{m} c^{A}_{lm}\,Y_l^m(\theta,\phi)+\epsilon P_5(\cos\theta)\right],\qquad \varepsilon\in\mathbb{R} \tag{S2}$$

where we have expanded $\mathbf{x}=(x,\theta,\phi)$ on the right, $P_5$ is the degree 5 Legendre Polynomial (i.e. $P_5(\cos\theta)\propto Y_5^0(\theta,\phi)$). By construction, $\rho(\mathbf{x})$ matches the tetrahedral point cloud at every $(l,m)\,\forall l\leq 4$. The term $\epsilon P_5(\cos\theta)$ can be understood as the orthogonal perturbation added on top of the tetrahedral density. By orthogonality of spherical harmonics, the perturbative term only contributes to $c_{5,0}$. Consequently, for a density expansion truncated at $l_{\max}=4$, the tetrahedral point cloud and the entire continuous family of surface densities are indistinguishable as they produce the same invariant descriptor $\boldsymbol{\xi}$. We can readily create more such examples by adding perturbative terms to Eq. (S2), which only contribute to $l>l_{\max}$.

This implies that inverting a finite set of expansion coefficients amounts to solving a constrained inverse problem that requires the recovered real space density to be representable as a discrete and finite point cloud. By treating atomic coordinates as the free parameters, instead of optimizing the expansion coefficients, the reconstruction procedure proposed in this work restricts the search space and ensures that every intermediate density expansion corresponds to a discrete point cloud.

* jnigam@mit.edu
† tsmidt@mit.edu

## II. COMPARING DESCRIPTOR IMPLEMENTATIONS

Several implementations of $\nu$-neighbor density correlations differ in how they construct and symmetrize the tensor product of density coefficients Eq. (4) (reproduced below)

$$\rho_i^{\otimes\nu} = \prod_{\alpha=1}^{\nu} c_i^{n_\alpha l_\alpha m_\alpha}.$$

In particular, two commuting group actions are defined on $\boldsymbol{\rho}_i^{\otimes\nu}$, namely

- the permutation group $S_\nu$, which acts by reordering $c_i^{l_\alpha m_\alpha}$ (or equivalently the tuple of $\nu$ indices $(l_{\alpha_1} m_{\alpha_1}, \ldots, l_{\alpha_\nu} m_{\alpha_\nu})$). $D(\pi)\boldsymbol{\rho}_i^{\otimes\nu} = \otimes_{\alpha=1}^{\nu} v_{\pi^{-1}(\alpha)}, \quad \pi \in S_\nu,$
- the rotation group $SO(3)$, which also acts through the tensor product representation $D(\hat{R}) = \bigotimes_{\alpha=1}^{\nu} D^{(l_\alpha)}(\hat{R}), \quad \hat{R} \in$ SO(3)

where $D(\pi)$ is the representation of the permutation group and $D^{(l_\alpha)}(\hat{R})$ is the Wigner-D matrix corresponding to the $(l_\alpha)$ irrep of rotation $\hat{R}$.

In the following, we briefly review how reduced tensor products (RTP) in e3nn, Gaunt tensor products, and the $N$-body iterative contraction of equivariants (NICE) enforce these symmetries.

### A. Reduced tensor products

RTP explicitly accounts for both of these symmetries by constructing a linear change of basis from the unsymmetrized tensor product $\rho_i^{\otimes\nu}$ to $\boldsymbol{\xi}^{\otimes\nu;\lambda\mu}$ whose components transform according to irreducible representations of both $S_\nu$ and SO(3). The change of basis is obtained by solving a linear intertwining equation that simultaneously enforces the two group actions [1]. In practice, this requires solving a linear system whose dimension grows rapidly with the correlation order $\nu$. One of the optimizations implemented in Ref. 1 is a multinomial decomposition of the symmetric tensor product of the density coefficients into a direct sum of the symmetric tensor products of the individual angular-momentum channels. In the following, we denote these features as $\boldsymbol{\xi}_{i,\mathrm{RTP}}$.

### B. NICE/ACDC

The recursive construction in NICE [2] takes the alternative approach of enforcing rotational symmetries through successive Clebsch-Gordan couplings of pairs of density coefficients (or, density coefficients and intermediate equivariant features), thereby avoiding a global change of basis. Permutation symmetry is accounted for by restricting the tuple of angular momenta $(l_1, \ldots, l_\nu)$ to a canonical ordering $l_1 \leq l_2 \leq \cdots \leq l_\nu$, so that tensor products related by permutations (forming different equivalence classes under permutations) need not be enumerated explicitly. As this construction underlies atom-centered density correlations (ACDC), we denote features as $\boldsymbol{\xi}_{i,\mathrm{NICE}}$ ad $\boldsymbol{\xi}_{i,\mathrm{ACDC}}$ interchangeably.

### C. Gaunt products

Unlike the previous methods, Gaunt products [3] compute equivariants through multiplicative products of spherical harmonics instead of tensor products of irreducible representations. The products are scaled by Gaunt coefficients

$$G_{l_1 m_1, l_2 m_2, l_3 m_3} = \int Y_{l_1}^{m_1}(\hat{\mathbf{r}}) Y_{l_2}^{m_2}(\hat{\mathbf{r}}) Y_{l_3}^{m_3}(\hat{\mathbf{r}})\, d\mathbf{r}, \tag{S3}$$

which are related to Clebsch-Gordan coefficients as

$$G_{l_1 m_1 l_2 m_2 l_3 m_3} = \sqrt{\frac{(2l_1+1)\,(2l_2+1)}{4\pi\,(2l_3+1)}}\, \langle l_1 0; l_2 0 | l_3 0\rangle\ \langle l_1 m_1; l_2 m_2 | l_3 m_3\rangle\,.$$

Products of spherical harmonics (which are functions on a 3D sphere) are mapped to a convolution in a 2D Fourier basis, which can then be efficiently computed using fast Fourier transforms. The selection rules of which features are

produced by the Gaunt product were characterized by [4]; in our setting here where we are looking at symmetric products, [4] confirms that all required components can be computed using the Gaunt product. As in NICE, higher-order features are computed recursively through pairwise convolutions, allowing the computation to follow the same canonical ordering of angular momenta. We denote these features as $\boldsymbol{\xi}_{i,\text{Gaunt}}$.

### D. Conventions and relationships between different implementations

The normalizations used in the implementations of RTP, Gaunt products, and ACDC differ only by simple multiplicative factors. These conversion factors depend only on the symmetries (angular momentum indices) of the interacting densities and the resulting feature symmetry ($\lambda$). In the following, we provide factors for translating descriptors between conventions. Each factor is expressed as a ratio and parameterized by the corresponding angular indices. For the equivariant power spectrum ($\nu = 2$), the features resulting from RTP and Gaunt products are related as

$$\frac{\boldsymbol{\xi}^{\otimes 2;\lambda\mu}_{i,\text{RTP}}}{\boldsymbol{\xi}^{\otimes 2;\lambda\mu}_{i,\text{Gaunt}}}(l_1, l_2, \lambda, l_{\max}) = \begin{cases} \frac{\sqrt{2(2l_{\max}+1)}}{\langle l_1 0; l_2 0|\lambda 0\rangle} & l_1 \neq l_2 \\ \frac{\sqrt{(2l_{\max}+1)}}{\langle l_1 0; l_2 0|\lambda 0\rangle} & l_1 = l_2 \end{cases}, \tag{S4}$$

and for the invariant bispectrum

$$\frac{\boldsymbol{\xi}^{\otimes 3}_{i,\text{RTP}}}{\boldsymbol{\xi}^{\otimes 3}_{i,\text{Gaunt}}}(l_1, l_2, l_3, l_{\max}) = \begin{cases} f(l_1, l_2, l_3, l_{\max}) & l_1 \neq l_2 \neq l_3 \\ \frac{f(l_1, l_2, l_3, l_{\max})}{\sqrt{6}} & l_1 = l_2 = l_3 \\ \frac{f(l_1, l_2, l_3, l_{\max})}{\sqrt{2}} & \text{otherwise} \end{cases}, \tag{S5}$$

where $f(l_1, l_2, l_3, l_{\max}) = (3l_{\max} + 1)\frac{\sqrt{6(2l_3+1)}}{\langle l_1 0; l_2 0|l_3 0\rangle}$.
The factors relating ACDC and RTP are even simpler and depend only on whether the resulting features are invariant ($\lambda = 0$) or equivariant ($\lambda > 0$),

$$\frac{\boldsymbol{\xi}^{\otimes 2;\lambda\mu}_{i,\text{RTP}}}{\boldsymbol{\xi}^{\otimes 2;\lambda\mu}_{i,\text{ACDC}}}(l_1, l_2, \lambda, l_{\max}) = \begin{cases} 1 & \lambda = 0 \\ \sqrt{2} & \lambda > 0 \end{cases} \tag{S6}$$

$$\frac{\boldsymbol{\xi}^{\otimes 3;\lambda\mu}_{i,\text{RTP}}}{\boldsymbol{\xi}^{\otimes 3;\lambda\mu}_{i,\text{ACDC}}}(l_1, l_2, \lambda, l_{\max}) = \begin{cases} 1 & \lambda = 0 \\ \sqrt{3} & \lambda > 0 \end{cases} \tag{S7}$$

Throughout this work, we present results using the ACDC implementation.

Finally, we note that the descriptors are compact for two reasons. First, we consider a single radial channel as detailed in the main text, and retain only the canonical ordering of angular momenta, $l_1 \leq l_2 \leq \cdots$ as discussed above. Second, we discard descriptors that vanish identically due to symmetry. Several such selection rules have been derived in Ref. 2. We describe an additional selection rule for $\nu = 3$ pseudoscalars below.

We first note that Clebsch-Gordan coefficients satisfy the exchange symmetry

$$\langle l_1 m_1; l_2 m_2|l_3 m_3\rangle = (-1)^{l_1+l_2-l_3}\langle l_2 m_2; l_1 m_1|l_3 m_3\rangle, \tag{S8}$$

which implies

$$\boldsymbol{\xi}^{\otimes 3;-1}_{i,(l_1,l_2,l_3)} = \sum_{m_1 m_2 m_3}\langle l_1 m_1; l_2 m_2|l_3 m_3\rangle \prod_{j=1}^{3} c_i^{l_j m_j} \tag{S9}$$

$$= (-1)^{l_1+l_2-l_3}\sum_{m_1 m_2 m_3}\langle l_2 m_2; l_1 m_1|l_3 m_3\rangle \prod_{j=1}^{3} c_i^{l_j m_j} \tag{S10}$$

In the case $l_1 = l_2$, relabeling $l_1 m_1$ as $l_2 m_2$ and vice-versa should not affect the sum,

$$\boldsymbol{\xi}^{\otimes 3;-1}_{i,(l_1,l_2,l_3)} = (-1)^{l_1+l_2-l_3}\boldsymbol{\xi}^{\otimes 3;-1}_{i,(l_2,l_1,l_3)} \tag{S11}$$

Using Eq. (S8) and $l_1 = l_2$,

$$\boldsymbol{\xi}^{\otimes 3;-1}_{i,(l_1,l_1,l_3)} = (-1)^{l_3} \boldsymbol{\xi}^{\otimes 3;-1}_{i,(l_1,l_1,l_3)} \tag{S12}$$

Now recall that $\boldsymbol{\xi}^{\otimes 3;-1}$ is a pseudoscalar, i.e., it accumulates a negative sign under inversion (i.e., $\sigma = -1$) (see SI of Ref. [2] for derivation). As $\sigma = (-1)^{l_1+l_2+l_3}$ by definition, when $l_1 = l_2$ a pseudoscalar must satisfy $-1 = (-1)^{l_3}$. Plugging into Eq. (S12), we get $\boldsymbol{\xi}^{\otimes 3;-1}_{i,(l_1,l_1,l_3)} = -1\boldsymbol{\xi}^{\otimes 3;-1}_{i,(l_1,l_1,l_3)}$, which implies

$$\boldsymbol{\xi}^{\otimes 3;-1}_{i,(l_1,l_1,l_3)} = 0.$$

## III. RECONSTRUCTION WITH AND WITHOUT PSEUDOSCALARS

A regular tetrahedron and its inverted (mirror) copy have identical scalar invariants ($\lambda = 0$, $\sigma = 1$), but differ in their pseudoscalar components ($\sigma = -1$), which change sign under inversion. When the pseudoscalar components are omitted, both structures are equally consistent with the target descriptor, and the reconstruction may converge to either structure. Including the pseudoscalars uniquely identifies the target configuration and improves reconstruction accuracy.

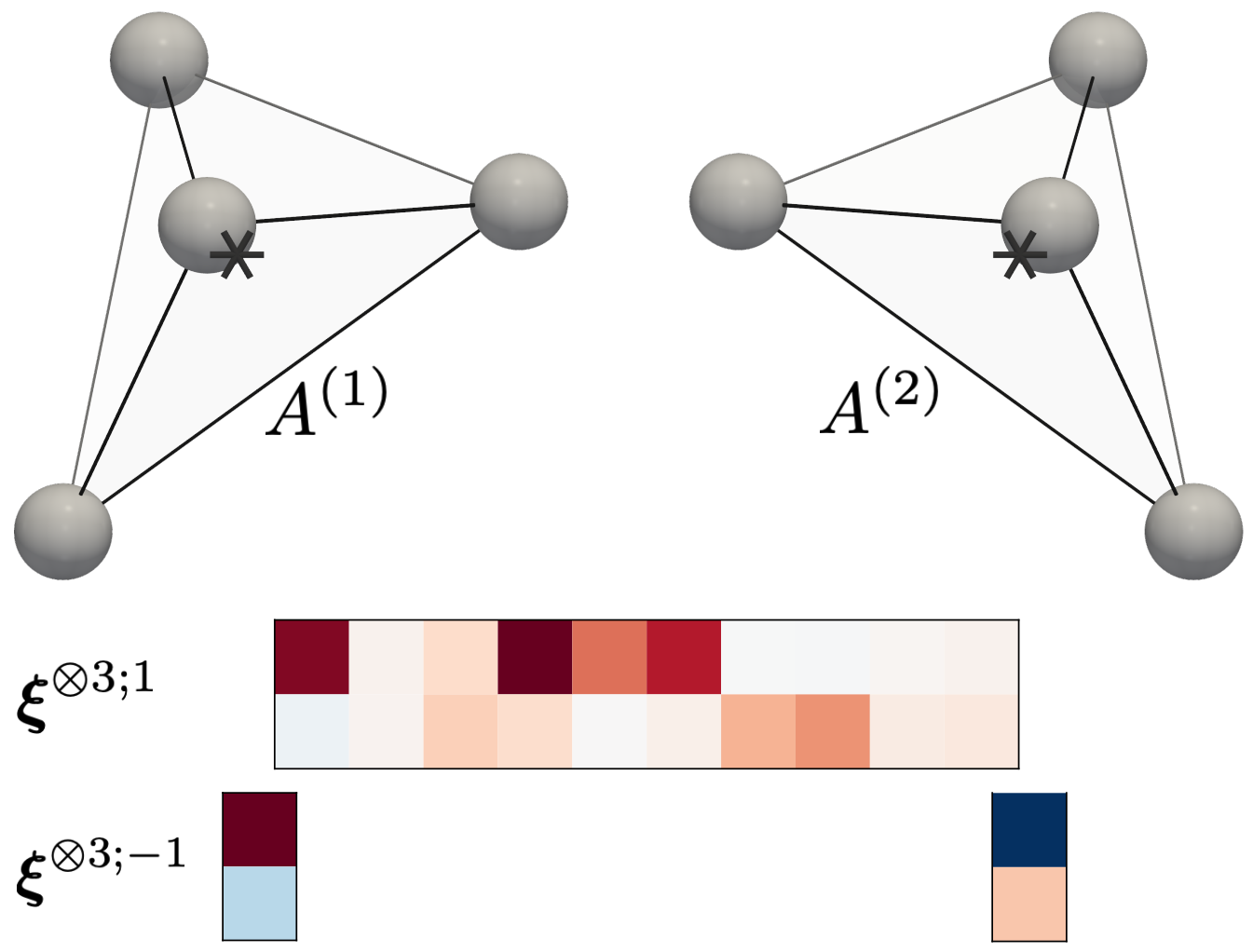


FIG. S1. $\nu = 3$ features computed with $l_{\max} = 5, n_{\max} = 1$, resulting in 20 scalar $\boldsymbol{\xi}^{\otimes 3,1}$ and two pseudoscalar $\boldsymbol{\xi}^{\otimes 3,-1}$ features for each environment. For both the right-handed ($A^{(1)}$) and left-handed ($A^{(2)}$) tetrahedra, the scalar features are identical, while the pseudoscalar features differ in sign, i.e., $\boldsymbol{\xi}^{\otimes 3;-1}(A^{(1)}) = -\boldsymbol{\xi}^{\otimes 3;-1}(A^{(2)})$.

## IV. EFFECT OF RADIAL BASIS

In the main text, we have described the projection of atom densities in Eq. (2) onto a discrete basis of radial $R_n(|\mathbf{x}|)$ and angular functions $Y_\ell^m(\hat{\mathbf{x}})$. The formulation in Eq. (3) reflects the choice of a single radial basis function $R_n(x) = x$. However, as noted in the main text, this choice can introduce infinite descriptor degeneracies. Along each direction $\hat{\mathbf{u}}$, all radial information is compressed into a single scalar, so configurations with different distributions of neighbor distances can have the same expansion coefficients. As this degeneracy affects $c_i^{lm}$, it propagates to all higher-order invariant features computed from these expansion coefficients (i.e., increasing angular resolution or correlation order does not resolve this degeneracy).

Most atomistic descriptors for machine learning instead introduce a set of radial basis functions $\{R_n\}_{n=1}^{n_{\max}}$. So each direction is specified by a vector $\left(\sum_{j=1}^{N_i} R_1(r_{ji}), \sum_{j=1}^{N_i} R_2(r_{ji}), \ldots, \sum_{j=1}^{N_i} R_{n_{\max}}(r_{ji})\right)$, which retains substantially more information about the radial distribution of neighbors. For the degeneracy to persist, two sets of neighbor distances $\{r_{ji}\}$ and $\{r'_{ji}\}$ that were indistinguishable in the single-basis limit must now simultaneously satisfy

$$\sum_j R_n(r_{ji}) = \sum_j R_n(r'_{ji}) \qquad \forall n \in \{1, \ldots, n_{\max}\}. \tag{S13}$$

When these radial basis spans polynomial-like functions (for instance, $R_n(x) = x^n$), this means that two configurations must have identical sequences of power sums $\{\sum_j r_{ji}^n\}_{n=1}^{n_{\max}}$, which is a stronger constraint than matching a single radial moment, and reduces the likelihood of an accidental degeneracy. The argument is equally applicable to other choices of radial bases, including Gaussian-type orbitals (GTOs), polynomials composed of $\{1, r, r^2, \dots\}$, etc.

In Sec. III C, the radial basis was chosen to be the orthonormal set of Jacobi polynomials $P_n^{(\alpha,\beta)}(x)$ with $\alpha, \beta > -1$.

### A. Constant single radial basis

On the other hand, one may choose a radial basis function which is identically constant for all inputs. Starting from the neighbor density

$$\rho_i(\mathbf{x}) = \sum_{j=1}^{N_i} \delta\left(\mathbf{x} - \mathbf{r}_{ji}\right),$$

and choosing a single constant radial basis function $R(x) = 1$, the projection of the atom density can be written as

$$c_i^{(lm)} = \int d\mathbf{x}\, \rho_i(\mathbf{x}) Y_l^m(\hat{\mathbf{x}}).$$

Considering only the invariant $l = 0$ coefficient, and substituting the density,

$$c_i^{(00)} = \int d\mathbf{x} \sum_j \delta\left(\mathbf{x} - \mathbf{r}_{ji}\right) \tag{S14}$$

$$= \sum_{j=1}^{N_i} 1 = N_i, \tag{S15}$$

which reduces to the number of neighbors in the environment (and contains no geometric information).

## V. COMPARING POINT CLOUDS

We quantify reconstruction accuracy using two complementary metrics. Since the optimization of Eq. (12) is performed over invariants, the recovered structure is determined only up to an arbitrary global rigid transformation, and in general, may not match the orientation of the reference environment.

The first metric avoids the need for a common frame of reference by comparing internal coordinates of the reference and reconstructed configurations, which are insensitive to rigid transformations. We compare two finite atomic configurations comprising $N_i$ neighbors (identical particles) through their distance matrices. The distance matrix [5, 6] $\mathbf{D}$ has elements of the form $\mathbf{D}_{i_\alpha i_\beta} = \|(\mathbf{r}_{i_\alpha i} - \mathbf{r}_{i_\beta i})\|$, where $\mathbf{r}_{i_\alpha i}$ are neighbor coordinates relative to the central atom $i$ (i.e., Euclidean interatomic distances). As all atoms are treated to be chemically identical in this work, two structures related only by a particle relabeling $\mathbf{P}$ (a permutation of row/column indices) are identical. This metric was also introduced as $d_{\text{ref}}$ in Ref. 7 and defined as,

$$d_{\text{ref}}(\mathbf{D}, \mathbf{D}') = \min_{\mathbf{P}} \|\mathbf{D} - \mathbf{P}\mathbf{D}'\mathbf{P}^\top\|,$$

where the minimum is taken over all permutations (an exhaustive search over all $N!$ permutations).

The second metric instead first aligns the reconstructed configuration with the reference structure using an optimal rigid transformation and then computes the root-mean-square deviation (RMSD) between the aligned atomic coordinates. The optimal alignment can be determined using the Kabsch (rigid Procrustes) algorithm (via SVD of the cross-covariance matrix) after centering both structures. Alternatively, instead of aligning the recovered and reference structures directly in Cartesian space, one can determine the optimal rotation of their corresponding expansion coefficients. Note that rotating a point cloud by $\hat{R} \in \mathrm{SO}(3)$, transforms the corresponding density expansion coefficients via the Wigner D-matrices,

$$c_i^{(lm)}(\hat{R}A) = D^l_{mm'}(\hat{R}) c_i^{(lm')}(A)$$

independently for each $l$.

Parameterizing rotations by unit quaternions $q \in S^3$, the optimal rotation is obtained by minimizing the mismatch

$$\mathcal{L}(q) = \|\mathbf{c}_i(\hat{R}(q)A) - \mathbf{c}_i(A)\|^2,$$

where the boldface $\mathbf{c}_i$ denotes the concatenated vector of expansion coefficients over all $l$ and $m$. We implement this numerically in PyTorch and perform the optimization of $\mathcal{L}(q)$ using the Adam optimizer.

## VI. EXPERIMENTAL DETAILS

In the following subsections we provide additional details and figures for the experiments reported in the main text.

### A. Recovering environments in molecular and material datasets

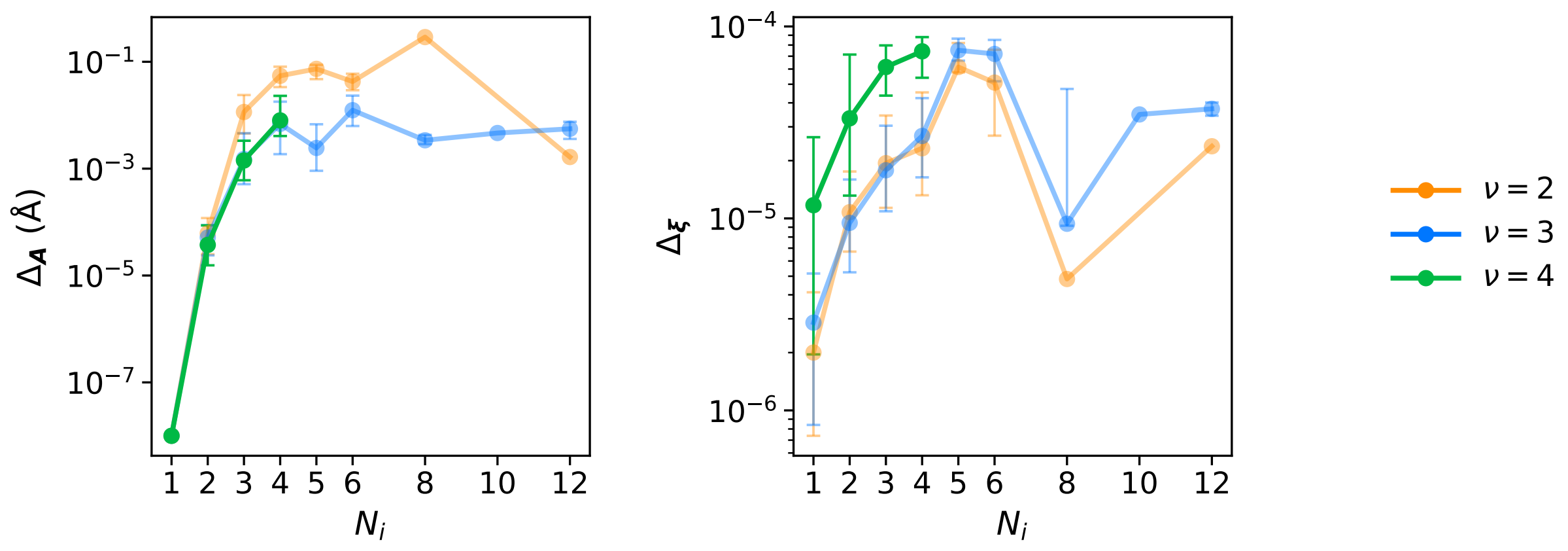


FIG. S2. Descriptor reconstruction error ($\Delta_{\boldsymbol{\xi}}$) and structural reconstruction error (RMSD, $\Delta_A$) as a function of coordination number for the power spectrum ($\nu = 2, l_{\max} = 8$), bispectrum ($\nu = 3, l_{\max} = 6$) and trispectrum ($\nu = 4, l_{\max} = 3$) for successful reconstructions. $\nu = 2, 3$ curves are reproduced from Fig. 2.

For reconstruction of atomic environments across subsets of QM9, MP20 and transition metal complexes described in Sec. III A, we report the reconstruction accuracy ($\Delta_{\boldsymbol{\xi}}$ and $\Delta_A$) for $\nu = 4, l_{\max} = 3$ in Fig. S2. The results for $\nu = 2$ and $\nu = 3$ are replotted from Fig. 2 for comparison. Although the $\nu = 4$ descriptor has a higher dimensionality (56) than the $\nu = 3$ or $\nu = 2$ descriptors, we obtain successful reconstructions only for 30% of the environments, showing that a larger descriptor dimensionality does not necessarily translate into an improved reconstruction. In particular, we do not obtain successful reconstruction (i.e., optimization loss or the MAE between target and optimized descriptors $\leq 10^{-5}$) for environments with $N_i > 4$ neighbors.

### B. Identifying near-accidental degeneracies

The atomic coordinates of the pair of environments optimized to be degenerate at $\nu = 4$ as described in Section III C of the main text are provided in Table VI B.

In Fig. S3, we show $\Delta_{\text{degen}}$ for an additional pair of environments, with 6 neighbors each, optimized to be nearly degenerate at $\nu = 3$. The resulting structures remain geometrically distinct, with $d_{\text{ref}} = 0.11$ Å. We displace the last atom by a distance $\pm\varepsilon$ along either (i) the direction obtained during the degeneracy optimization or (ii) a randomly chosen direction. In both cases, the descriptor difference, $\Delta_{\text{degen}}$ is computed between the positively and negatively displaced structures. When the displacement follows the optimized direction (left), the $\nu = 2$ and $\nu = 3$ descriptor differences remain close to zero over the entire range of $\varepsilon$, indicating that the descriptor is nearly insensitive to perturbations along this direction. The degeneracy is lifted, and the structures are distinguishable by a higher-order descriptor of correlation order $\nu = 4$. In contrast, when the same atom is instead displaced by $\pm\varepsilon$ along a random direction (right), the descriptor differences increase rapidly with $\varepsilon$ for all correlation orders. Hence, the

| $A_i^{+\varepsilon}$ | | | $A_i^{-\varepsilon}$ | | |
|---|---|---|---|---|---|
| $x$ | $y$ | $z$ | $x$ | $y$ | $z$ |
| 0.0000 | 0.0000 | 0.0000 | 0.0000 | 0.0000 | 0.0000 |
| -0.355 | 0.6752 | 1.0378 | -0.3551 | 0.6752 | 1.0378 |
| -1.024 | 0.4752 | -0.6205 | -1.0240 | 0.4752 | -0.6205 |
| 0.0907 | -0.7137 | -0.0079 | 0.0907 | -0.7137 | -0.0079 |
| 0.2960 | 1.9819 | 0.1200 | 0.2960 | 1.9819 | 0.1200 |
| 1.5549 | -0.2578 | 1.2433 | 1.5549 | -0.2578 | 1.2433 |
| 0.6795 | 0.0711 | -0.2251 | 0.6795 | 0.0711 | -0.2251 |
| -0.1441 | -0.4506 | 0.7656 | 0.1441 | 0.4506 | -0.7656 |

TABLE I. Cartesian coordinates (Å) of the pair of $\nu = 4$ degenerate structures shown in Fig. 5. Note that only the position of the last atom differs between the two structures. The two structures are geometrically distinct, with $d_{\mathrm{ref}} = 2.092$ Å.

near-degeneracy is associated with a specific direction in configuration space rather than a general lack of sensitivity of the descriptors.

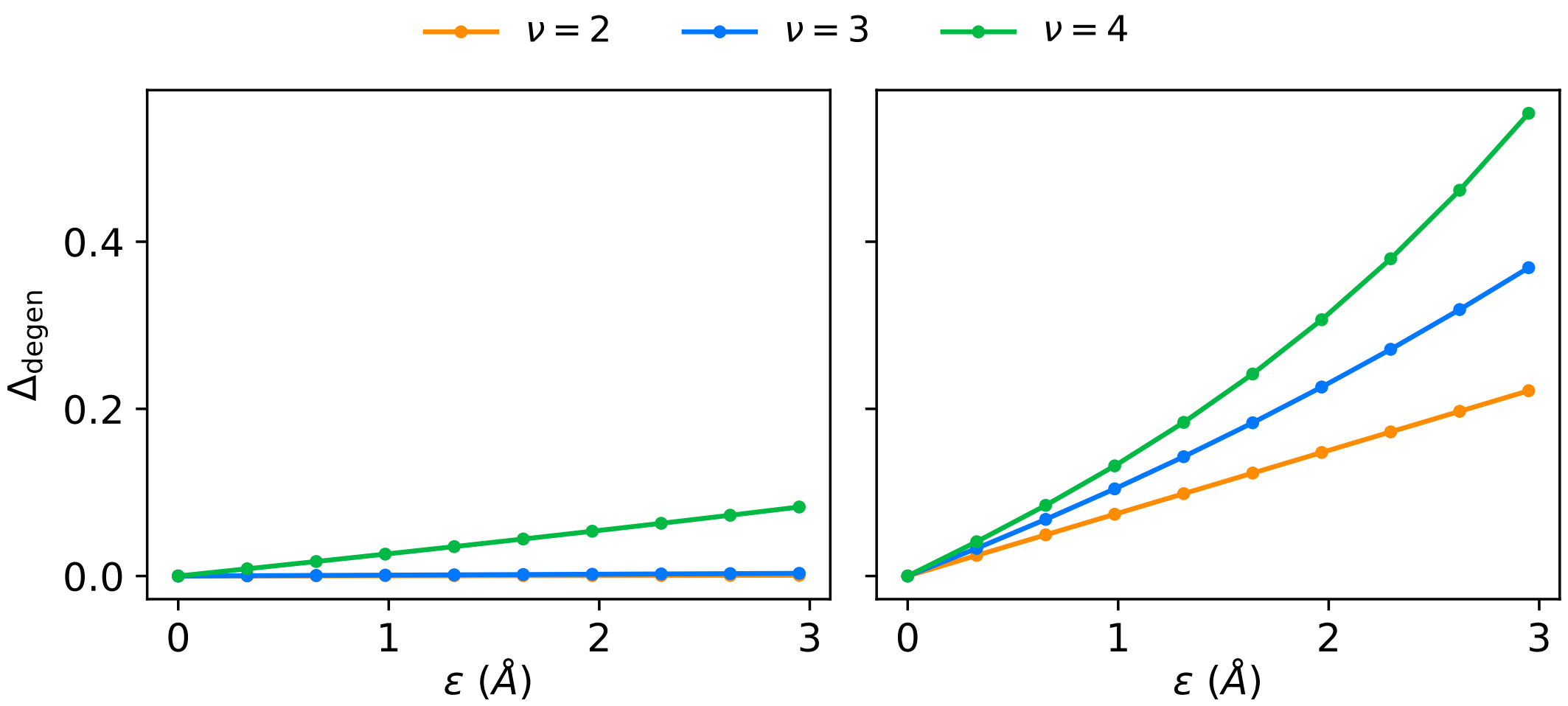


FIG. S3. Difference between descriptors $\Delta_{\mathrm{degen}} = \boldsymbol{\xi}(A_i^+) - \boldsymbol{\xi}(A_i^-)$ for a pair of nearly degenerate environments as a function of displacement magnitude $\varepsilon$ of the perturbed atom for $\nu = 2$ (orange), $\nu = 3$ (blue) and $\nu = 4$ (green). (Left) The atom is displaced along the direction optimized for $\nu = 3$ degeneracy (meaning to minimize $\Delta_{\mathrm{degen}}$ for $\nu = 3$). The optimized pair of structures corresponds to $\varepsilon = 0.8$ AA. (Right) For the same range of displacement magnitudes but along a random direction, $\Delta_{\mathrm{degen}}$ grows substantially faster for all three correlation orders, confirming that the near-degeneracy on the left is specific to the optimized direction.

### C. Interpolating descriptors corresponding to the Bailar twist

Although the descriptors corresponding to the actual Bailar twist trajectory ($\boldsymbol{\xi}(A_{i,\phi})$) and the linear interpolation trajectory ($\boldsymbol{\xi}_\tau$) appear to be qualitatively similar in Fig. 6 of the main text, their components differ significantly, as shown in Fig. S4. The top panel shows the component-wise bispectrum mismatch between the two trajectories, while the bottom panel highlights the descriptor component with the largest deviation along the path.

#### *1. Optimizing to environments with six neighbors*

In the main text, we observed that the linear interpolation in descriptor space frequently reconstructs to structures with fewer than six neighbors. Here, we repeat the reconstruction while constraining every optimized environment to comprise six neighbors. As shown in Fig. S5, this causes the descriptor error $\Delta_{\boldsymbol{\xi}}$ to increase compared to the optimization that did not constrain the number of neighbors. The recovered structures also incur RMSDs $\Delta_A$ of

| $A_i^{+\varepsilon}$ | | | $A_i^{-\varepsilon}$ | | |
|---|---|---|---|---|---|
| $x$ | $y$ | $z$ | $x$ | $y$ | $z$ |
| 0.0000 | 0.0000 | 0.0000 | 0.0000 | 0.0000 | 0.0000 |
| -0.5476 | 1.5416 | 1.3603 | -0.5476 | 1.5416 | 1.3603 |
| -1.3810 | 0.7602 | -0.8331 | -1.3810 | 0.7602 | -0.8331 |
| -0.1331 | -0.8678 | -0.1182 | -0.1331 | -0.8678 | -0.1182 |
| 0.3068 | 2.1057 | 0.2739 | 0.3068 | 2.1057 | 0.2739 |
| 1.3641 | 0.0548 | 1.1392 | 1.3641 | 0.0548 | 1.1392 |
| 0.6395 | -0.1610 | 0.4529 | -0.6395 | 0.1610 | -0.4529 |

TABLE II. Cartesian coordinates (Å) of the pair of $\nu = 3$ degenerate structures shown in Fig. S3. Note that only the position of the last atom differs between the two structures. The two structures are geometrically distinct, with $d_{\text{ref}} = 0.11$ Å.

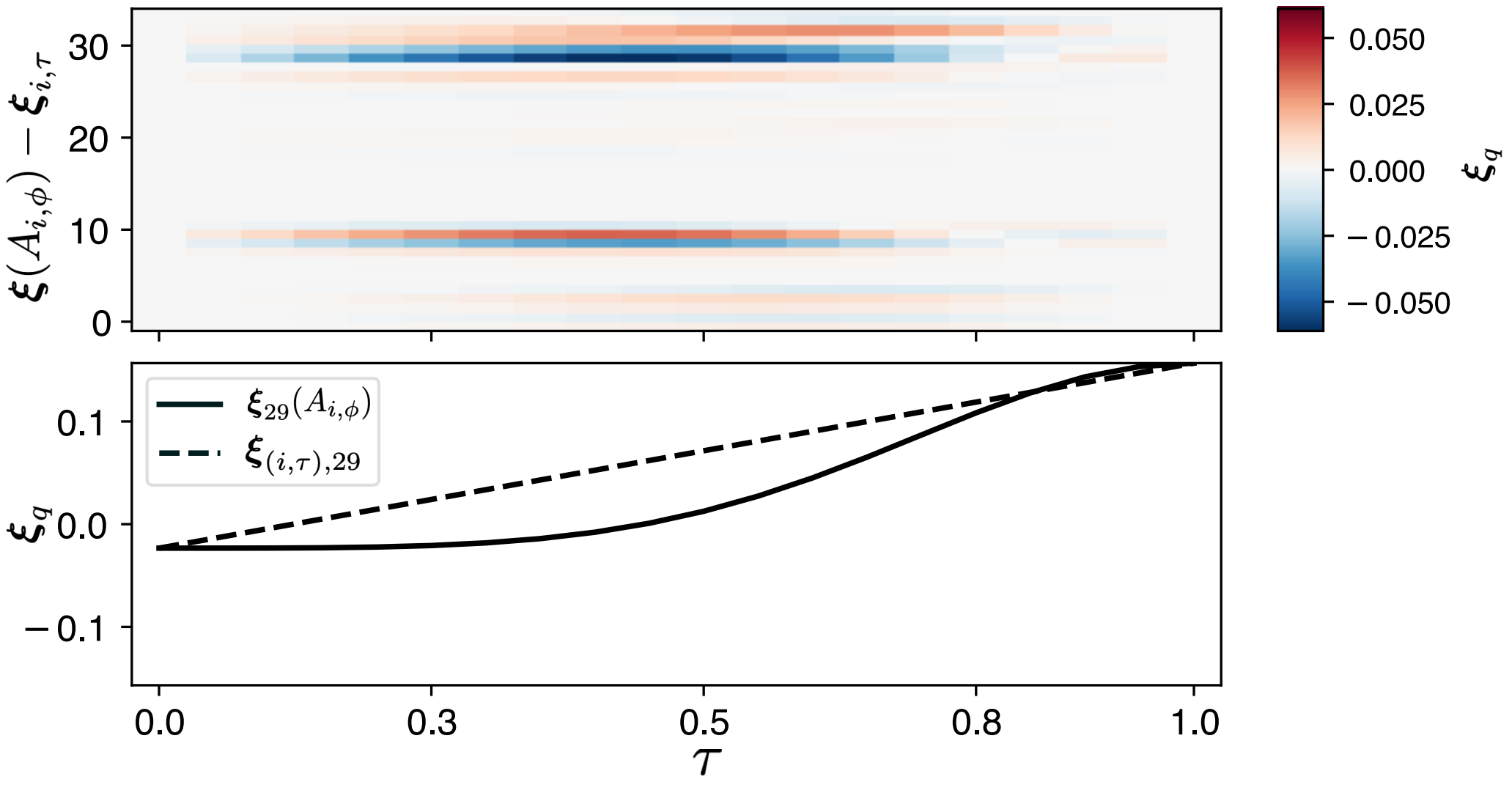


FIG. S4. Descriptor mismatch $\boldsymbol{\xi}(A_{i,\phi}) - \boldsymbol{\xi}(A_{i,\tau})$ for $\nu = 3$ correlations along the interpolation path. The bottom panel compares the descriptor corresponding to the reference trajectory (the Bailar twist) (solid) and linear interpolation (dashed) for feature index $q = 29$, the component with the largest peak $\|\Delta\boldsymbol{\xi}_q\|$ along the path.

up to 0.15 Å with respect to the geometric twist. Unlike the Bailar twist, which preserves the geometry of the two triangular faces and only changes their relative twist angle, the optimized structures suffer a substantial distortion of the triangular faces (visualized in Fig. S6).

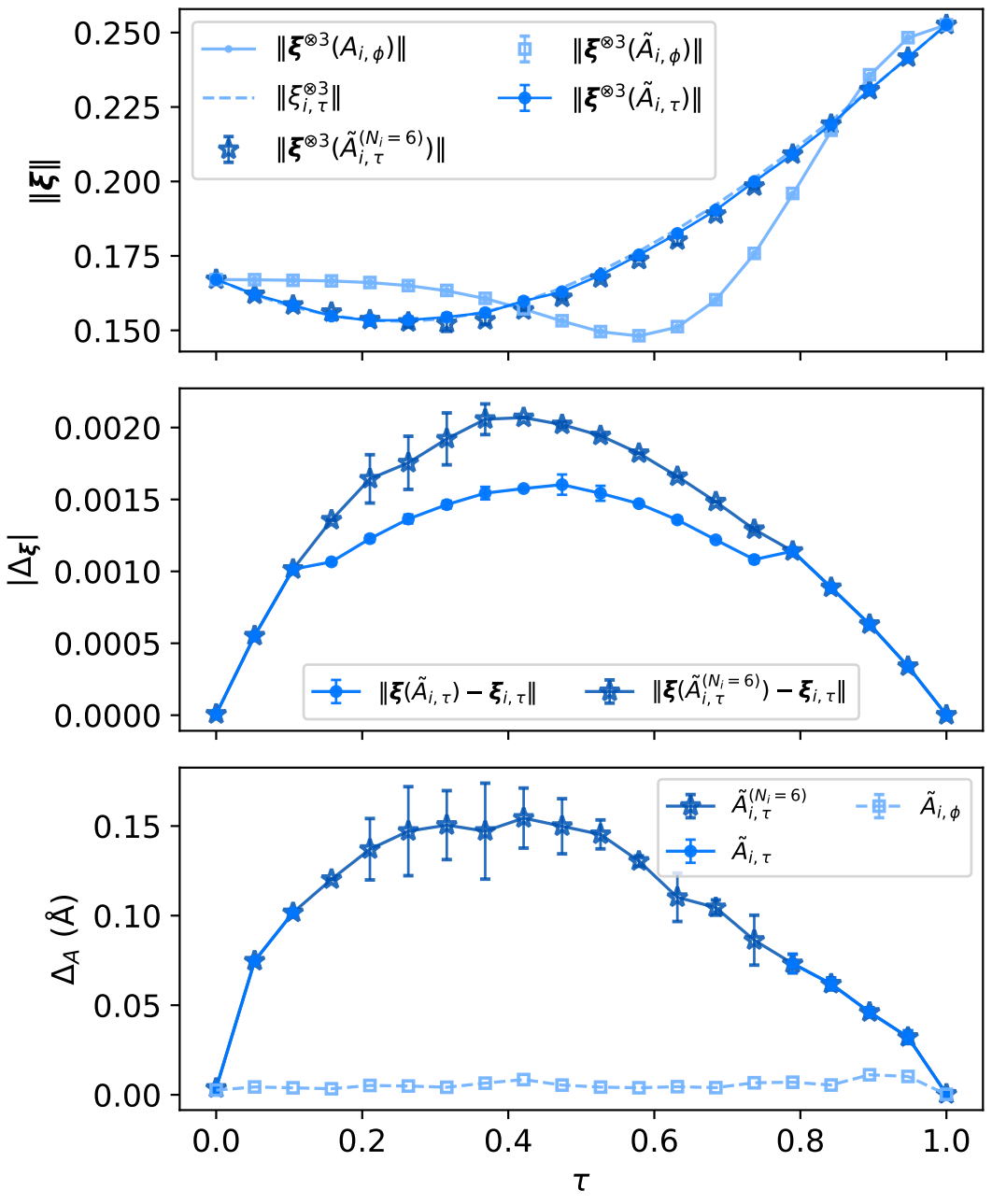


FIG. S5. Companion to Fig. 6, showing the effect of constraining the number of neighbors during reconstruction. As before, solid lines and filled circles correspond to the reference geometric twist $A_{i,\phi}$ and its reconstruction $\tilde{A}_{i,\phi}$, dashed lines and open squares correspond to the linearly interpolated descriptor $\boldsymbol{\xi}_{i,\tau}$ and reconstructed structures $\tilde{A}_{i,\tau}$. We use the star marker to show the analogous reconstructed quantities (descriptor norms, errors, and RMSD) when the number of neighbors is constrained to $N_i = 6$ throughout the trajectory, denoted as $\tilde{A}_{i,\tau}^{(N_i=6)}$. The constrained reconstructions exhibit a larger descriptor mismatch at intermediate interpolation points and RMSDs of up to 0.15 Å compared to the true Bailar trajectory.

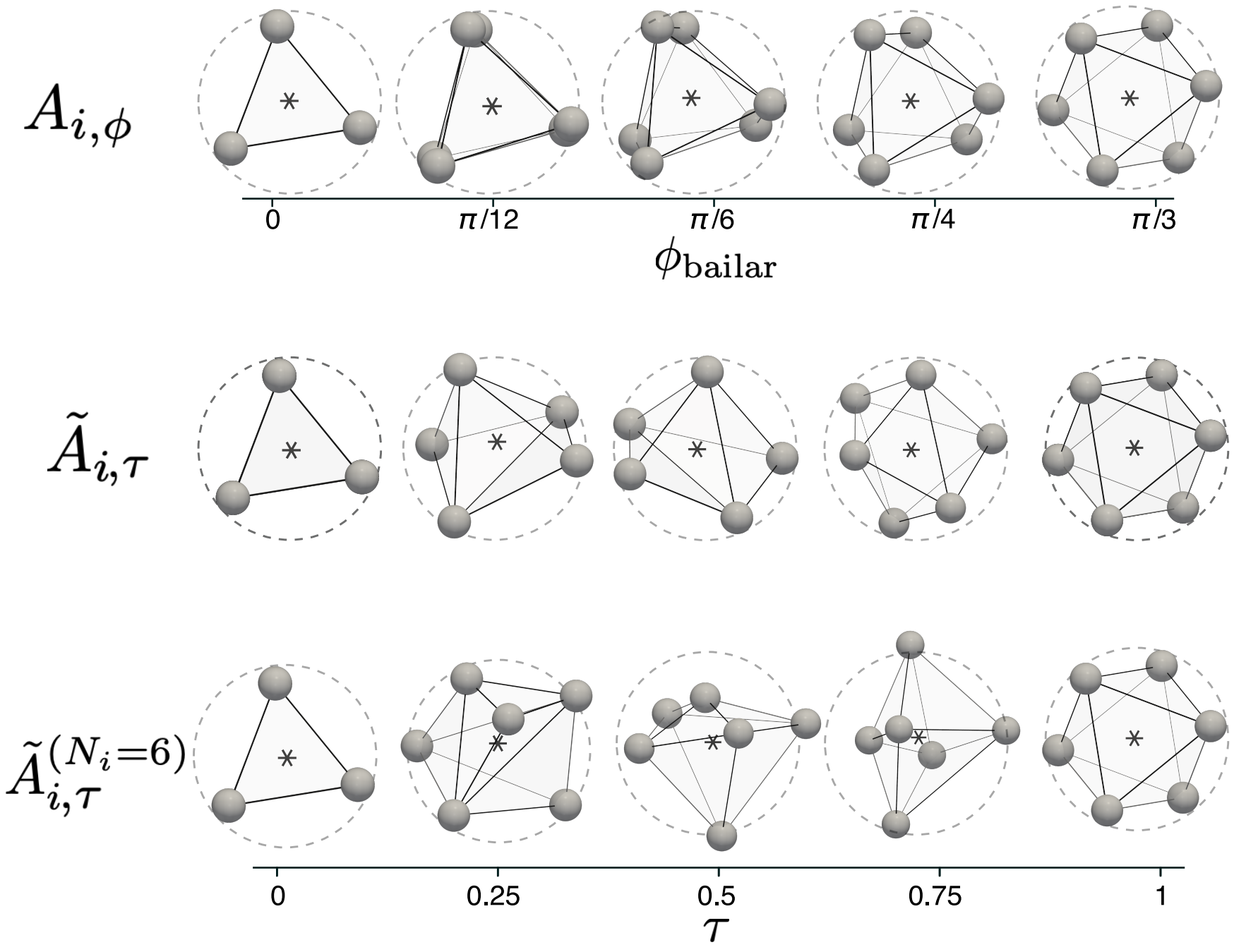


FIG. S6. Five snapshots of the representative geometries along the Bailar-twist interpolation. $A_{i,\phi}$ (top row) are the true reference structures along the Bailar twist, from the trigonal prism ($\phi = 0$) to the octahedron $\phi = \pi/3$. Environments reconstructed from the linearly interpolated descriptor $\boldsymbol{\xi}_{i,\tau}$, $\tilde{A}_{i,\tau}$ (middle row) with the number of neighbors left unconstrained. Structures reconstructed from the same interpolated descriptors, but with the coordination number constrained to $N_i = 6$ (bottom row)



\end{document}